\documentclass[letterpaper,twocolumn,10pt]{article}
\PassOptionsToPackage{hyphens}{url}
\usepackage{usenix}

\usepackage{xcolor}

\usepackage{amsmath,amssymb,amsfonts}
\usepackage{amsthm}

\newtheorem{theorem}{Theorem}[section]

\theoremstyle{definition}

\theoremstyle{remark}

\usepackage{booktabs}
\usepackage{makecell}
\usepackage{multirow}
\usepackage{multicol}
\usepackage{graphicx}
\usepackage{subcaption}

\usepackage{enumitem}
\usepackage{placeins}

\usepackage[ruled,vlined]{algorithm2e}

\SetKwInOut{KwIn}{Input}
\SetKwInOut{KwOut}{Output}

\newlength{\BreakAlgoMargin}


\begin{document}

\date{}

\title{
  \Large\bf
  Towards TEE-Certified DP:\\
  Verifiable Differentially Private Training on Legacy GPUs
}

\author{
Li Ge\textsuperscript{1,*},
Wenjie Qu\textsuperscript{2,*},
Weitao Feng\textsuperscript{1},
Yi Zeng\textsuperscript{1},
Jiaheng Zhang\textsuperscript{2},
Xiaofeng Wang\textsuperscript{1},
Wei Dong\textsuperscript{1}
\\[1mm]
\textsuperscript{1}Nanyang Technological University
\qquad
\textsuperscript{2}National University of Singapore
\\
\textsuperscript{*}Equal contribution.
}

\maketitle

\begin{abstract}

Wide adoption of machine learning has created growing policy and
regulatory demand for protecting sensitive training data, with
differential privacy (DP) emerging as a key mechanism. Yet a
less-studied problem is how to certify the faithful execution of DP
during training: an external verifier should be able to check that a
released model was trained with proper DP protection, without accessing
the private training data. Existing cryptographic approaches, such as
zero-knowledge proofs, provide strong guarantees but often incur
prohibitive overhead, in some cases by orders of magnitude. Trusted
Execution Environments (TEEs) offer a more efficient alternative, but
the multi-GPU TEE support needed for training and fine-tuning large
language models remains limited to recent platforms and is absent or
inefficient on legacy GPUs.

To address this, we propose a practical framework for
verifiable DP training using CPU-side TEEs together with untrusted
GPUs. Our design addresses a fundamental efficiency-security tension:
training entirely inside a CPU TEE is too slow, while unrestricted GPU
offloading can allow malicious deviations from DP. We therefore offload
expensive gradient computation to GPUs, while using the CPU TEE to
efficiently verify the correct enforcement of DP on gradients through
probabilistic checking. Our framework detects frequent full deviations from DP with high
probability; for the utility-oriented forged-gradient attacks evaluated in
this work, sparse deviations provide limited utility benefit and show no
measurable additional membership leakage. Experiments further show that our approach nearly
achieves a ``free lunch'': it incurs only modest overhead compared with
standard GPU-based DP training, while effectively constraining
malicious deviations from the claimed DP execution.
\end{abstract}

\section{Introduction}

As sensitive data such as electronic health records and tax records are increasingly used in machine learning (ML) training, privacy risks, including reidentification, membership inference, and attribute exposure, have become a serious concern~\cite{fredrikson2015model,carlini2022membership}. The dominant mitigation is \textit{Differential Privacy} (DP)~\cite{dwork2006calibrating}, which ensures that any single record has only a limited effect on an algorithm’s output. In ML training, DP is commonly implemented through \textit{DP-SGD}~\cite{abadi2016deep}, which adds calibrated noise to per-example gradients before aggregation, so the final model satisfies DP. DP and DP-SGD have thus been increasingly recommended by privacy laws, regulations, and best-practice guidelines for model protection~\cite{NISTDP2023,OECDPET2023,abowd2018us}.

\vspace{2pt}\noindent\textbf{Verifiable DP training}. However, assuring that DP-SGD is faithfully enforced during model training remains challenging. The key issue is verifying the \emph{integrity} of the training process: whether the declared DP mechanism has been executed correctly throughout training without deviation. This is critical because the noise injected at each iteration inevitably degrades model utility, creating incentives for model owners to deviate from the protocol while still claiming privacy protection. For example, a company subject to regulatory requirements or public commitments may reduce or even omit DP noise to obtain a higher-performing model.
Importantly, this concern is not merely hypothetical. Several real-world deployments of DP have attracted scrutiny regarding whether their implementations provide the same guarantees as claimed~\cite{tang2017privacy}. A recent work reports that one building block in Apple's DP framework in real-time is configured with 
DP disabled, thus uploading data without any DP protection~\cite{chourasia2026auditing}. 
Although extensive research has focused on designing DP mechanisms~\cite{dwork2014algorithmic,mironov2017renyi} and analyzing or auditing their privacy guarantees~\cite{jagielski2020auditing}, little attention has been paid to verifying that DP-SGD is actually executed as claimed, despite the importance of such assurance for privacy protection and regulatory compliance.

Addressing this concern requires \textit{verifiable DP}~\cite{narayan2015verifiable,biswas2022verifiable}, which aims to provide \emph{publicly checkable evidence} that a claimed DP mechanism has been executed as specified. Given such evidence, together with the declared protocol and the resulting model, an independent verifier should be able to determine that the model was produced under the attested verification protocol and its declared probabilistic security bounds. Importantly, this verification itself must preserve privacy and should not require access to sensitive training data. In the example above, after a company trains a model with a claimed DP guarantee, a regulator or external auditor may require evidence showing that the declared DP mechanism was actually enforced during training.

Verifiable privacy mechanisms have been studied for tasks such as $k$-means clustering~\cite{narayan2015verifiable}, DP counting, and histogram release~\cite{narayan2015verifiable,biswas2022verifiable}. Extending such guarantees to deep model training, however, remains challenging. Existing state-of-the-art approaches~\cite {shahin2024confidential} rely on \textit{zero-knowledge proofs} (ZKPs) to certify the correctness of the training procedure. While providing strong guarantees, ZKPs incur substantial computational and communication costs that grow with model size and the number of training iterations. Even for relatively small models with only $0.01$B parameters, the overhead can exceed $1000\times$, making these approaches impractical for realistic deep learning workloads.

\vspace{2pt}\noindent\textbf{TEE-based alternative}. Beyond heavyweight ZKP-based solutions, a natural alternative is to use
hardware-based \textit{trusted execution environments} (TEEs). A TEE provides an
isolated execution environment whose code identity and execution state can be
remotely attested. This suggests a simple design: run the entire training
procedure inside a TEE and use the hardware root of trust to certify, through an
open-source monitor, that the DP protocol was correctly executed. An external
verifier can then check both that the declared DP protocol was faithfully
enforced and that the released model is %
bound to the attested
execution.

However, this ``train-inside-a-TEE'' design faces a practical obstacle:
training reasonably sized models requires GPUs. Full confidential-computing
support across both CPU and GPU TEEs is available only on the most recent
platforms, such as NVIDIA Blackwell B200~\cite{nvidia2024nvidia}. Older GPUs, including H100
and H200 based on the Hopper architecture, rely on secure I/O through encrypted
bounce buffers, which introduces substantial training overhead~\cite{apsey2024confidential,zhu2024confidential}.
Earlier-generation GPUs, such as V100, A100, and RTX 4090, provide no TEE
support at all. Since GPU infrastructure is costly and replaced slowly, it is
unrealistic to assume that GPU-TEEs will be widely available in the near
future~\cite{wang2025game}.

\vspace{2pt}\noindent\textbf{Our approach}. We therefore propose a new approach: combining widely deployed CPU-TEEs
with untrusted GPUs. CPU-TEEs, such as AMD SEV~\cite{sev2020strengthening} and Intel TDX~\cite{aktas2023intel},
are already available on commodity servers, but are too slow to run full
training by themselves (Section~\ref{section:experiment}). Our key idea is instead to use the
CPU-TEE to \emph{monitor} DP-SGD executed on untrusted GPUs and generate
verifiable evidence of protocol compliance. This design preserves the
efficiency of GPU-based training while providing verifiable DP guarantees with
only modest overhead over conventional, non-verifiable DP-SGD.

More specifically, we design and implement a CPU-TEE-based framework for verifiable DP training that incurs only modest overhead compared with standard GPU-based DP-SGD. In the absence of GPU TEE support and efficient secure I/O, our framework achieves high performance by offloading expensive gradient computation to untrusted local GPUs while keeping trusted randomness, optimizer and model-state maintenance, and verification inside the CPU TEE; the GPU performs per-example gradient computation, clipping, and aggregation, subject to probabilistic trusted recomputation. This architecture, however, introduces a  challenge: an untrusted trainer may tamper with GPU-computed gradients before they enter the TEE, causing the resulting model to deviate from the claimed DP protocol (Section~\ref{section:challenge}).

To address this challenge, we classify malicious deviations according to their frequency, distinguishing between \emph{dense} and \emph{sparse} deviations. To understand how much utility a trainer can obtain from infrequent manipulation, we study several utility-oriented forged-gradient attacks that steer the DP trajectory toward non-private checkpoints. Across the attack family and workloads evaluated in this paper, small attack budgets provide limited utility gain and no measurable additional membership leakage, leaving a rational trainer with limited incentive to employ them; we use these experiments to select an empirical sparse--dense operating threshold rather than to claim that every possible sparse attack is ineffective. Motivated by this observation, we design an efficient CPU-TEE-based protocol that probabilistically detects repeated full deviations while explicitly accounting for the cumulative manipulation admitted through its numerical-tolerance region. The GPU performs per-example gradient computation, clipping, and aggregation, whereas the CPU-TEE holds the trusted randomness and training state, generates the DP noise, and probabilistically verifies the GPU-side computation. As a result, frequent deviations are detected with high probability, while the impact of infrequent deviations remains bounded by DP-SGD clipping, enabling efficient verifiable DP training under realistic hardware assumptions. %

\vspace{2pt}\noindent\textbf{Contributions}. Our contributions are listed below:

\vspace{2pt}\noindent$\bullet$\textit{~System Design:} 
  We propose a CPU-TEE-based solution for verifiable DP training that is compatible with legacy GPU infrastructure. We identify an inherent efficiency--security tension in this setting. To address this tension, we design a framework that achieves a practical trade-off between security and efficiency. Our framework offloads gradient computation, per-example clipping, and aggregation to untrusted GPUs, while keeping trusted randomness, model and optimizer state, DP-noise generation, and probabilistic verification inside the CPU-side TEE. The protocol probabilistically detects repeated full deviations while explicitly accounting for the cumulative manipulation admitted through its numerical-tolerance region, and avoids the cost of fully verifying every training step.
  
\vspace{2pt}\noindent$\bullet$\textit{~Adversarial Analysis:} We study three utility-oriented checkpoint-steering attacks---Current, Nearest, and Final---under adversarially selected attack schedules. Across the evaluated workloads, attack budgets below $\overline{M}=50$ malicious iterations ($0.33\%$--$0.89\%$ of the training steps on the LLM workloads, $1\%$ on CIFAR-10) provide limited or no utility improvement, and our membership-inference evaluation shows no measurable additional leakage over the honest DP baselines. These findings motivate $\overline{M}=50$ as an empirical operating threshold for the attack family considered here; they do not claim to exhaust all possible utility- or privacy-oriented attacks.

\vspace{2pt}\noindent$\bullet$\textit{~Numerical Discrepancy as an Attack Surface:} We identify that the numerical discrepancy between honest GPU and TEE executions is itself an attack surface: any tolerance the verifier must grant for honest disagreement can be exploited by a dishonest GPU to hide bounded manipulations that no single check would flag. Independently of the attack strategy, our two-stage verification protocol classifies every committed submission into a body, ambiguity, or hard-rejection region, meters tolerated discrepancies observed on checked steps with bounded ledgers, and yields an analytic high-probability security budget for the cumulative normalized deviation that can pass through the first two regions, alongside a Bernoulli detection guarantee for the third.

\vspace{2pt}\noindent$\bullet$\textit{~Experimental Results:} 
  We conduct a comprehensive empirical evaluation across multiple models, datasets, and training configurations. 
  Our experiments show that verifiable DP training is nearly a ``free lunch'' on LLM oriented tasks: our framework incurs only $4.4\%$--$14.9\%$ overhead compared with standard GPU-based DP training across a range of tasks and datasets. In contrast, executing the full training procedure inside a CPU-side TEE incurs $8.0\times$--$9.2\times$ overhead.

\section{Preliminaries}

\subsection{Differential Privacy}
Differential Privacy (DP) \cite{dwork2006calibrating} is a rigorous framework for quantifying the privacy protection provided by a randomized algorithm. Intuitively, DP requires that the output of an algorithm be only minimally affected by the presence or absence of any individual data record, thereby limiting the information that can be inferred about a particular participant.

In this work, we adopt the add/remove-one notion of neighboring datasets. Two datasets $\mathcal{D}$ and $\mathcal{D}'$ are said to be neighboring, denoted by $\mathcal{D} \sim \mathcal{D}'$, if they differ by the addition or removal of a single record. Given privacy parameters $\varepsilon,\delta \ge 0$, a randomized algorithm $\mathcal{M}$ is said to satisfy $(\varepsilon,\delta)$-DP if for any pair of neighboring datasets $\mathcal{D} \sim \mathcal{D}'$ and any measurable set of outputs $\mathcal{S}$,
\begin{align*}
\Pr[\mathcal{M}(\mathcal{D}) \in \mathcal{S}]
\leq
e^{\varepsilon}
\Pr[\mathcal{M}(\mathcal{D}') \in \mathcal{S}]
+
\delta .
\end{align*}
Therefore, DP requires the output distributions of $\mathcal{M}$ on neighboring datasets to be close, thereby limiting privacy leakage. In the context of private model training, $\mathcal{M}(\mathcal{D})$ can be viewed as the model produced by the randomized training algorithm on dataset $\mathcal{D}$.

\subsection{DP-SGD}
Differentially Private Stochastic Gradient Descent (DP-SGD)~\cite{abadi2016deep} enforces differential privacy during training by bounding the influence of each individual example through \emph{per-example gradient clipping} and injecting \emph{Gaussian noise} into the aggregated gradient. Clipping prevents any single record from dominating the update, while the added noise masks the contribution of individual examples and provides a formal privacy guarantee.

Let $D=\{x_i\}_{i=1}^{N}$ be the training dataset, $\boldsymbol{\theta}$ the model parameters, and $\ell(\boldsymbol{\theta};x)$ the loss for an example $x$. At training iteration $t$, given clipping norm $C$, noise multiplier $\sigma$, and learning rate $\eta_t$, the per-example gradient on $x_i$ is
\begin{align*}
\boldsymbol{g}_t(x_i)
=
\nabla_{\boldsymbol{\theta}}\ell(\boldsymbol{\theta}_t;x_i).
\end{align*}
Each gradient is clipped to have $\ell_2$ norm at most $C$:
\begin{align*}
\bar{\boldsymbol{g}}_t(x_i)
=
\boldsymbol{g}_t(x_i)
\cdot
\min\left(
1,
\frac{C}{\|\boldsymbol{g}_t(x_i)\|_2}
\right).
\end{align*}
DP-SGD then aggregates the clipped gradients in the training batch $B_t$ and adds Gaussian noise:
\begin{align*}
\widetilde{\boldsymbol{g}}_t
=
\frac{1}{|B_t|}
\left(
\sum_{x_i\in B_t}
\bar{\boldsymbol{g}}_t(x_i)
+
\boldsymbol{z}_t
\right),
\qquad
\boldsymbol{z}_t
\sim
\mathcal{N}
\left(
\boldsymbol{0},
\sigma^2 C^2\boldsymbol{I}
\right).
\end{align*}
Finally, the model parameters are updated as
\begin{align*}
\boldsymbol{\theta}_{t+1}
=
\boldsymbol{\theta}_t
-
\eta_t
\widetilde{\boldsymbol{g}}_t.
\end{align*}

\subsection{Trusted Execution Environment}

A Trusted Execution Environment (TEE) is a hardware-backed isolated execution context that provides \emph{confidentiality} and \emph{integrity} for code and data during execution, even if the host operating system or hypervisor is compromised. Typical TEEs (e.g., Intel SGX, AMD SEV-style enclaves) offer attestation mechanisms: a remote party can verify that a specific program (identified by a measurement such as a hash) is running inside a genuine TEE before trusting its outputs. In our setting, attestation enables an auditor to trust that the DP-enforcing logic (noise generation, model updates, and the recomputation that checks GPU-side clipping and aggregation) is executed as specified.

However, due to legacy deployment constraints discussed in the introduction, most widely available TEEs today are CPU-side TEEs, which impose practical limitations.
Enclave transitions can be expensive, and full-scale training entirely within a TEE is often impractical—especially for large models that rely on high-throughput GPU computation. Therefore, we treat the TEE as a \emph{trusted core} that holds the trusted randomness, the optimizer and model state, and the verification logic, and records verifiable evidence, while outsourcing heavy gradient computation to an untrusted but fast GPU worker whose results are checked by probabilistic trusted recomputation.  Unless otherwise specified, all TEEs are referred to as CPU-TEEs throughout this document.

\section{Problem Definition and Security Model}
In this work, we consider a three-party setting that captures the practical deployment of verifiable DP training.

\vspace{2pt}\noindent\textbf{User (data provider).}
The \emph{user} contributes data records, such as samples and labels, to form the training dataset. Users care about privacy:
they require a formal DP guarantee that bounds each participant's privacy
leakage and limits what can be inferred about their individual records from
the released model. 

\vspace{2pt}\noindent\textbf{Company (trainer).}
The \emph{trainer} (e.g., a company operating the training pipeline and GPU workers) controls the model architecture, training pipeline, and computational resources. The trainer aims to obtain a high-utility model efficiently. However, because differential privacy typically reduces model utility, a profit-driven or malicious trainer may have incentives to deviate from the prescribed DP-SGD protocol while still claiming compliance to users or regulators. 

In particular, the trainer may attempt to improve model utility by modifying key components of DP-SGD. Typical deviations include: (i) skipping or weakening gradient clipping, thereby increasing sensitivity; (ii) reducing the noise scale $\sigma$ or replacing Gaussian noise with a weaker distribution; (iii) reusing noise across steps or employing predictable randomness; (iv) altering the sampling procedure, for example by using larger effective batch sizes than declared; and (v) selectively applying DP mechanisms only to a subset of training steps. 

\vspace{2pt}\noindent\textbf{Government (regulator/auditor).}
The \emph{government} (or regulator) is an external party that enforces compliance with privacy requirements. The regulator does not necessarily participate in training, but it demands \emph{verifiability}: the company should provide convincing evidence that DP-SGD was executed correctly with the claimed hyperparameters (e.g., sampling rate, clipping norm, and noise multiplier). 

\vspace{2pt}\noindent\textbf{Interaction among the three parties.}
The interaction proceeds as follows. Users provide data to the company, expecting that the model will be trained under the advertised DP guarantee. The company then trains the model using its own infrastructure, which may include both untrusted high-performance components, such as GPUs, and trusted components, such as a CPU-based TEE. During training, the company is expected to follow the declared DP-SGD protocol and produce evidence that the regulator can later use to assess compliance.

The regulator does not need to observe every low-level operation. Instead, it requires that privacy-critical parts of the protocol be either protected by trusted hardware or auditable afterward. In this sense, the TEE serves as a bridge between the company and the regulator: it protects key DP operations and helps generate verifiable evidence that the claimed protocol was enforced. Users therefore rely on the regulator’s compliance framework rather than directly observing the training process.

Our goal is to achieve verifiable DP training, where the training procedure generates a post-training certificate that can be released to users or used for future attestation. Such a design constrains the company from deviating from the prescribed protocol. As a result, the company is expected to faithfully follow the DP-SGD procedure, users obtain meaningful privacy protection, and regulators or users receive verifiable evidence rather than relying on unaudited privacy claims. 

\section{Challenges of Verifiable DP Training}
\label{section:challenge}
In this section, we identify the challenges of realizing verifiable DP training under realistic hardware assumptions.
We begin by considering two natural approaches and show that each is limited by either efficiency or security concerns.
Together, these limitations reveal a fundamental efficiency-security tradeoff that motivates our design.

\medskip

\noindent\textbf{TEE-only training.}
A straightforward way to achieve verifiable DP training is to execute the entire training process inside a TEE. Under this design, model parameters, gradient computation, clipping, noise generation, and parameter updates are all performed within the trusted boundary. Through remote attestation, a verifier can therefore obtain strong assurance that the declared DP-SGD protocol has been faithfully executed.

\textit{Efficiency limitation.} 
Although conceptually simple, this approach is impractical for modern learning workloads. Large-scale training, such as training a large language model, relies heavily on GPU acceleration, whereas CPU-based TEEs provide substantially lower computational throughput. As a result, repeatedly performing forward and backward propagation inside a CPU-TEE incurs prohibitive overhead. As shown in Section~\ref{section:experiment}, this design slows training by approximately $8.0\times$--$9.2\times$ on the evaluated LLM workloads and by about $20\times$ on CIFAR-10 compared with standard GPU-based DP training. Consequently, fully executing DP-SGD inside a CPU-TEE is too expensive for realistic deployments.

\smallskip \noindent\textbf{GPU-TEE split training.}
To improve efficiency, a natural alternative is to offload computationally intensive operations from the trusted environment to untrusted local hardware. 
In this design, the GPU performs the expensive forward and backward passes, while the TEE maintains the trusted model state and executes the privacy-preserving update. 
Concretely, at each iteration, the GPU computes the gradients for the current mini-batch and sends them to the TEE. The TEE then performs the DP-SGD update, including clipping and noise addition, updates the trusted model parameters, and then returns the updated state information to the local GPU for the next iteration.

\textit{Security limitation.}
While substantially more efficient, this design introduces a critical security gap. The TEE only observes the gradients received from the GPU and cannot directly verify whether they were produced by honest forward and backward propagation on the intended model and training batch. Consequently, a malicious trainer may manipulate GPU-side computation while continuing to interact with the TEE in a seemingly legitimate manner.
More concretely, the GPU may return gradients that do not correspond to the claimed training process. By carefully crafting such gradients, the trainer can steer optimization toward a higher-utility model while violating the integrity of the declared DP-SGD execution. For example, the trainer may maintain two models in parallel: a DP-compliant model that remains consistent with the TEE's view and a non-private model trained outside the trusted environment. During training, manipulated gradients can gradually reduce the discrepancy between these two models, allowing the final model to benefit from non-private training while still appearing to follow the declared DP protocol.

\smallskip \noindent\textbf{Security--efficiency tradeoff.}
The above discussion reveals an efficiency-security tradeoff in verifiable DP training. Executing the entire training procedure inside a TEE provides strong integrity of DP guarantees but incurs prohibitive computational overhead. In contrast, offloading computation to untrusted GPUs achieves practical efficiency but creates opportunities for malicious deviations that undermine the integrity of the claimed DP execution.
This raises a central challenge: how can we retain the efficiency of GPU-accelerated training while still providing strong, verifiable guarantee that the declared DP-SGD procedure has been faithfully executed?

\section{Security--Efficiency Balance} %
In this section, we present our solution for verifiable DP training, which aims to achieve both practical efficiency and strong security. 
Our high-level idea is to tolerate sparse malicious deviations, which for the utility-oriented attacks we evaluate provide little utility benefit and no measurable additional membership leakage, while efficiently detecting dense malicious deviations that could meaningfully compromise the declared DP training execution.
The section proceeds as follows. First, we introduce a frequency-based classification of malicious deviations and distinguish between dense and sparse attack regimes (Section~\ref{section:classifying_malicious_activities}).
We then analyze the impact of sparse deviations on model utility and privacy leakage (Section~\ref{section:analysis_on_malicious_behavior}): for the utility-oriented attack family evaluated in this work, sufficiently small attack budgets provide limited utility benefit and no measurable additional membership leakage.

\subsection{Sparse and Dense Deviations}
\label{section:classifying_malicious_activities}

As discussed in the previous section, verifiable DP training exhibits an inherent tension between security and efficiency. The two straightforward solutions represent opposite ends of this tradeoff. Executing the entire training procedure inside a TEE can, in principle, eliminate adversarial behavior, but incurs substantial computational overhead. In contrast, offloading computationally intensive operations to an untrusted GPU greatly improves efficiency, but sacrifices the ability to verify computations performed outside the trusted boundary.
Therefore, a natural direction is to seek an intermediate design that balances security and efficiency. 
Instead of requiring the protocol to rule out all possible adversarial behaviors, we tolerate certain adversarial behaviors under a controlled relaxation of the security guarantee. Such tolerance must satisfy three properties:
\begin{itemize}[leftmargin=*]
    \item \textit{Negligible benefit from tolerated adversarial behaviors.}
    Since a malicious trainer is primarily motivated by improving model utility, tolerated deviations should provide little or no utility advantage over honest execution. If deviating from the prescribed protocol yields negligible benefit, then the trainer's incentive to cheat is substantially reduced.
    \item \textit{No significant harm to privacy protection.}
    Tolerated attacks should not introduce significant additional privacy leakage.
    \item \textit{Significant efficiency improvement.}
    The relaxation should enable substantial efficiency gains relative to fully trusted training, making verifiable DP training practical in realistic deployments.
\end{itemize}
The first two properties bound the incentive and potential harm of tolerated deviations, while the third ensures that the relaxation yields meaningful efficiency gains. 
Together, these conditions suggest that a favorable balance between security and efficiency can be achieved by tolerating certain low-impact adversarial behaviors.

To identify such behaviors, we perform a finer-grained analysis of malicious deviations.
Intuitively, some deviations can significantly alter the training trajectory, providing substantial utility gains to the trainer and potentially increasing privacy leakage.
Other deviations affect only a small number of training steps. For the utility-oriented attacks evaluated in this work, we observe that such sparse deviations have only limited influence on final model utility and no measurable additional membership leakage.

A natural way to distinguish between these two cases is by the number of malicious iterations during training.
Let $M$ be the total number of iterations on which the trainer deviates from the prescribed protocol, and let $\overline{M}$ be the maximum number of malicious iterations that the protocol is willing to tolerate. We classify malicious deviations below:
\begin{itemize}[topsep=3pt,itemsep=2pt,parsep=0pt,partopsep=0pt]
    \item \textit{Dense malicious activities.}
    If $M \ge \overline{M}$, we refer to the malicious deviations as dense.
    \item \textit{Sparse malicious activities.}
    If $M < \overline{M}$, we refer to the malicious deviations as sparse.
\end{itemize}

This distinction captures an important asymmetry between impact and detectability.
Dense deviations can substantially influence the optimization trajectory and therefore may provide meaningful utility gains to the trainer.
However, because they occur on many iterations, they are also easier to detect through probabilistic verification.
In contrast, sparse deviations are more difficult to detect because they occur on only a small number of iterations and would require substantially more verification effort to catch reliably.
As we show later for the utility-oriented attacks we evaluate, sparse deviations offer only limited utility improvement and no measurable additional membership leakage.

These observations motivate using deviations below the empirical threshold $\overline{M}$ as the tolerated operating regime in our system evaluation.
However, two challenges remain.
First, how should $\overline{M}$ be chosen to distinguish sparse attacks from dense ones in practice?
Second, what are the concrete utility benefits and privacy harms introduced by deviations below this threshold?
Later, we answer these questions through a comprehensive analysis of how the number of malicious iterations in DP training affects utility gain and privacy loss.

\subsection{Analysis of Sparse Malicious Deviations}
\label{section:analysis_on_malicious_behavior}

Now, we further analyze the effect of $M$ on model utility enhancement and privacy leakage through both theoretical analysis and empirical evaluation.

\subsubsection{Malicious Deviation Strategies}
\label{section:malicious_deviation_strategies}

We first characterize the malicious deviation strategies considered in our analysis.
As discussed earlier, a malicious trainer can steer the final DP model toward the corresponding non-private model by manipulating gradients during a subset of training iterations. Fundamentally, this attack aims to deviate the DP training trajectory toward a non-DP trajectory.
Since a training trajectory can be viewed as a sequence of checkpoints, the adversary may locally maintain checkpoints from non-private training. Then, at each malicious iteration of DP training, the adversary selects a target non-DP checkpoint and manipulates the submitted gradient to move the TEE-held model toward that checkpoint. Below, we study three natural and straightforward strategies for selecting this target checkpoint.

\begin{itemize}[topsep=3pt,itemsep=2pt,parsep=0pt,partopsep=0pt]
\item \textit{Current-checkpoint attack.} 
The most straightforward strategy is to select the checkpoint corresponding to the current training step. Specifically, at the $t$-th training iteration, the adversary selects the look-ahead checkpoint along the non-private trajectory as the target.
\item \textit{Nearest-checkpoint attack.} The adversary selects the non-private checkpoint closest to the TEE-held model.
\item \textit{Final-checkpoint attack.} Throughout the training process, the adversary can always fix the target checkpoint to the final checkpoint of the non-private model.
\end{itemize}

\begin{figure*}
    \centering
    \includegraphics[width=0.9\textwidth]{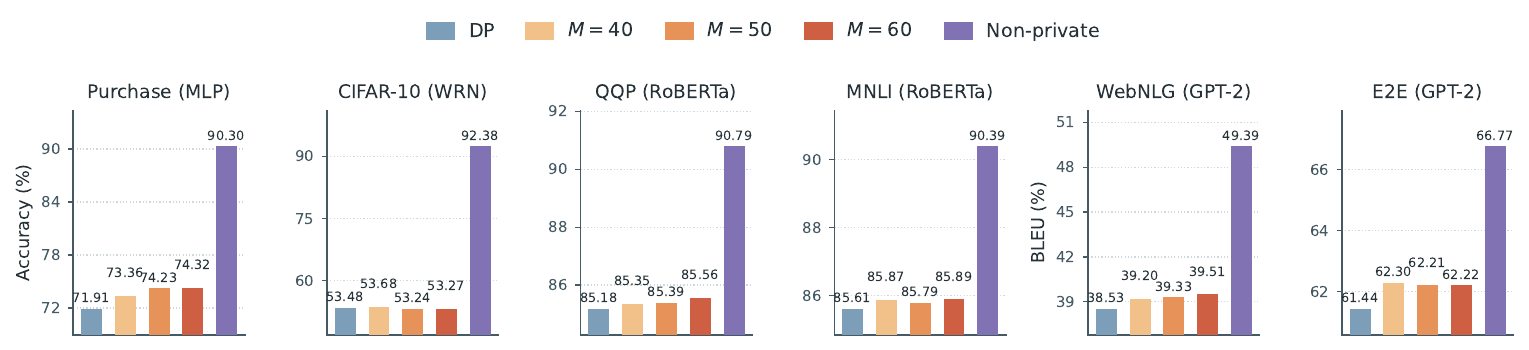}
    \caption{Utility comparison under sparse malicious deviations for $\varepsilon=2.0$. The two RoBERTa panels use RoBERTa-large and the two GPT-2 panels GPT-2-medium. For each task, $M=40,50,60$ report the best result among Current, Nearest, and Final attacks. Purchase, CIFAR-10, and QQP use accuracy, MNLI uses averaged matched/mismatched accuracy, E2E and WebNLG use BLEU, with the decoding configuration selected per bar on the validation split, so the bars within a panel do not share one configuration. The remaining metrics the two scorers emit are reported in Appendix~\ref{appendix:aux-metrics}.}
    \label{fig:grouped_bar}
\end{figure*}

Under any of the above strategies the adversary also chooses \emph{which} $M$ iterations to deviate on, and we grant it the strongest choice rather than a random one. Both schedules concentrate on $u^\star$, the fraction of training at which one unit of deviation buys the most displacement at the end ($u^\star=0.20$ for RoBERTa, $0.10$ for GPT-2): the final-checkpoint attack takes the $M$ consecutive steps centred on $u^\star T$, while the current- and nearest-checkpoint attacks take one step from each window of width $T/M$, at the point of that window nearest $u^\star T$. We did not run a uniform-random control, so these numbers should be read as the attacker's best placement rather than as an average over placements.
Given a selected target checkpoint, the details of how the adversary constructs the manipulated gradient to move the current model toward the checkpoint are provided in Appendix~\ref{appendix:attack-adamw}; the per-setting constants and the remaining configuration are listed in Appendix~\ref{appendix:attack-config} and Table~\ref{tab:attack-params}.

\subsubsection{Utility Analysis}
After characterizing the malicious deviation strategies, we now examine how the number of malicious iterations, $M$, affects the utility gain achievable by an adversary.
We use this attack procedure to evaluate the utility benefit obtainable under different numbers of malicious iterations, focusing on $M \in \{40,50,60\}$ in this section. Additional experiments with a broader range of $M$ values are reported in Section~\ref{section:experiment}. We conduct experiments on both conventional learning tasks and LLM-oriented fine-tuning tasks. For conventional learning, we consider a 5-layer MLP on Purchase and WideResNet on CIFAR-10. For LLM-oriented fine-tuning, we fine-tune GPT-2-medium on E2E and WebNLG and RoBERTa-large on QQP and MNLI.\footnote{We use LoRA for fine-tuning, a common parameter-efficient approach for reducing the overhead of DP training.} We evaluate malicious DP training and honest DP training under privacy budgets $\varepsilon = 2.0$, along with non-private training, whose checkpoints serve as the attacker's reference trajectory.

Figure~\ref{fig:grouped_bar} summarizes the utility impact of sparse malicious deviations under $\varepsilon=2.0$. For each task and each value of $M$, we report the best performance achieved among the three attack strategies, thereby representing the strongest utility gain obtainable by the adversary in our evaluation.
The results show that deviations with at most $50$ malicious iterations lead to only minor utility improvements over honest DP training in the experimental setting, especially when compared with the much larger utility gap between honest DP training and non-DP training. More precisely, when $M=50$, all tasks except Purchase improve by less than one point over the DP baseline; even on Purchase, where the improvement is the most noticeable, the gain is only about $2.3$ points.
For CIFAR-10 the deviations produce no consistent direction at all: the three budgets move accuracy by $+0.20$, $-0.24$ and $-0.21$ points, less than a quarter of a point either way.
We therefore set $\overline{M}=50$ as the sparse--dense deviation threshold; this corresponds to $0.33\%$--$0.89\%$ of the training steps on the LLM workloads and $1\%$ on CIFAR-10 (Table~\ref{tab:attack-params}).%
\footnote{The attack budget $M$ counts manipulated iterations; the full-deviation count $M_{\mathrm{full}}$ of Section~\ref{section:security_analysis} counts iterations that a check would reject. For this construction the two coincide unless the honest aggregate nearly equals the forged one: the forged aggregate lies on the clipping-ball boundary, so an attacked iteration falls inside the tolerance region only if $\lVert g^h\rVert_2\ge(1-\rho_{\mathrm{amb}})C\approx0.99\,C$, whereas the largest honest clipped average we measured was $0.65\,C$ on the GPT-2 tasks and $0.38\,C$ on the RoBERTa tasks. We treat the attacked iterations as full deviations on this basis, without re-verifying each one against the trusted reference.} Among the strategies evaluated here, none substantially closes the utility gap between honest DP and non-private training in this regime. We do not claim that the Current, Nearest, and Final strategies exhaust all possible utility-improving attacks: $\overline{M}=50$ is an empirical operating point supported by the strongest attacks in the evaluated family, and the protocol-level analyses of Sections~\ref{section:gpu-tee-discrepancy} and~\ref{section:security_analysis} are stated independently of this family.

\vspace{2pt}\noindent\textbf{Connection to bounded adversarial gradient perturbations.}
This assumption is consistent with a broader principle in optimization: when adversarial gradient perturbations are bounded in magnitude or frequency, their effect on the final model is also bounded. Recent theoretical work~\cite{yufei2026ijcai} shows that, for convex and smooth objectives, well-bounded gradient perturbations do not cause the learning process to deviate significantly; we discuss this connection in Appendix~\ref{appendix:bounded_gradient_perturbation}.

\begin{theorem}[Privacy of DP Training with Incomplete Verification]
\label{thm:privacy_incomplete_training}
Consider a $T$-iteration DP training procedure over a dataset of size $N$.
Suppose the trainer deviates from the prescribed protocol on $M$ iterations
and, in the worst case, each deviating iteration may submit an arbitrary
data-dependent vector of norm at most $C$, while the TEE still applies the
prescribed Gaussian noise. If the original training procedure is
$(\varepsilon,\delta)$-DP under the add/remove-one neighboring relation, then,
under the standard small-sampling-rate Gaussian-accounting approximation, the
resulting procedure is approximately $(\varepsilon',\delta)$-DP, where
\begin{align}
\frac{\varepsilon'}{\varepsilon}
\approx
\sqrt{
1+\frac{M}{T}
\left(
4N^2-1
\right)
}.
\end{align}
\end{theorem}

\noindent\textbf{Does this theorem really imply a significant privacy loss?}
Theorem~\ref{thm:privacy_incomplete_training} implies a worst-case
amplification of the privacy loss by a factor of approximately
$2N\sqrt{M/T}$ when the malicious term dominates, which can indeed reach
hundreds in practical settings. However, this seemingly large factor arises
from an extremely pessimistic attack that is quite different from the
utility-oriented deviations considered in our threat model. Specifically,
the $4N^2$ term allows the submitted vector on neighboring datasets to move
between opposite points of the clipping ball, giving sensitivity up to $2C$,
and allows this worst-case dependence to concentrate repeatedly on the same
target user. Repeating such highly targeted updates may maximize the privacy
leakage of one particular user, but it provides little reason to expect a
corresponding improvement in the overall utility of the trained model. As a
result, a rational utility-oriented adversary has little incentive to conduct
such attacks.

In contrast, utility-oriented deviations such as the common ones discussed
in Section~\ref{section:malicious_deviation_strategies} rely on broader, globally useful training signals rather
than repeatedly concentrating the updates on a single user's information.
For illustration, suppose a manipulated gradient averages $\kappa$ clipped
user contributions in such a way that its per-user sensitivity is at most
$2C/\kappa$. Under the same approximation,
\begin{align}
\frac{\varepsilon'}{\varepsilon}
\approx
\sqrt{
1+\frac{M}{T}
\left(
\frac{4N^2}{\kappa^2}-1
\right)
}.
\end{align}
Thus, when $\kappa$ is on the same order as $N$ and $M/T$ is small, the
resulting privacy inflation can be close to one. This calculation is
illustrative: dependence on many users alone does not imply the
$1/\kappa$ sensitivity reduction; the latter requires the manipulated signal
to average their contributions with correspondingly bounded per-user
influence. To empirically validate this theoretical intuition, we next
evaluate membership inference attacks and find that, for the utility-oriented
sparse deviations evaluated ($M\le 60$), there is no measurable additional
membership leakage beyond the variation of the honest DP baselines.

\vspace{2pt}\noindent
\textbf{Empirical study.} To quantify the actual privacy leakage introduced by sparse malicious deviations, we evaluate two membership inference attacks: IMIA~\cite{du2025imitative} and SHAPOOL~\cite{bai2026toward}, following the primary metric used in each attack's original evaluation protocol. 
The detailed attack configurations are provided in Appendix~\ref{appendix:mia_parameters}.
As shown in Table~\ref{tab:mia_purchase_cifar10_combined}, across both datasets Purchase and CIFAR-10, sparse deviations with $M=40$-$60$ yield MIA performance very close to the corresponding honest DP baselines and substantially below that of non-private training.
For SHAPOOL, the results are particularly stable: sparse malicious deviations achieve AUC values identical to, or within $0.01$ of, those of the corresponding honest DP baselines, indicating no measurable additional membership leakage.
For IMIA, the results exhibit greater variance. Nevertheless, the score values for both honest DP training and training with sparse deviations remain far below those of the corresponding non-private models. Interestingly, increasing either $\varepsilon$ or $M$ does not consistently increase the measured leakage. This non-monotonic behavior suggests that the stochastic variation inherent in DP training and MIA evaluation can be comparable to, or even larger than, the additional effect introduced by sparse malicious deviations.
These results do not rule out sparse attacks designed to maximize privacy leakage: Theorem~\ref{thm:privacy_incomplete_training} characterizes a substantially worse case in which malicious iterations repeatedly expose one target user, whereas the attacks evaluated here use broad training signals.

Overall, for the utility-oriented sparse attacks studied here, we observe limited utility improvement and no measurable additional membership leakage. These findings motivate tolerating a small number of deviations from an incentive perspective; they are not a universal guarantee for arbitrary sparse strategies, and the protocol analysis that follows classifies arbitrary submitted gradients by their trusted verification outcome rather than relying on these attack constructions.
The remaining task is therefore to design an efficient protocol that probabilistically detects repeated full deviations and accounts for the deviations admitted through numerical tolerance.

\begin{table}[t]
\centering
\caption{MIA performances on Purchase and CIFAR-10. DP stands for the DP-trained model, and $M=40, 50, 60$ stands for the malicious deviated models. IMIA and SHAPOOL are two MIA methods, and TPR and AUC are corresponding metrics.}
\label{tab:mia_purchase_cifar10_combined}
\scriptsize
\setlength{\tabcolsep}{2.5pt}
\renewcommand{\arraystretch}{0.88}
\begin{tabular}{ccccc ccc}
\toprule
\multirow{2}{*}{$\varepsilon$} 
& \multirow{2}{*}{Setting} 
& \multicolumn{3}{c}{\textbf{IMIA: TPR@0.1\% FPR}} 
& \multicolumn{3}{c}{\textbf{SHAPOOL: AUC}} \\
\cmidrule(lr){3-5} \cmidrule(lr){6-8}
& & Current & Nearest & Final & Current & Nearest & Final \\
\midrule
\multicolumn{8}{c}{\textbf{Purchase}} \\
\midrule
Non-Private & -- 
& \multicolumn{3}{c}{0.43} 
& \multicolumn{3}{c}{0.62} \\
\midrule
\multirow{4}{*}{$2.0$} 
& DP & 0.12 & 0.12 & 0.12 & 0.52 & 0.52 & 0.52 \\
& $M=40$ & 0.08 & 0.10 & 0.10 & 0.52 & 0.52 & 0.52 \\
& $M=50$ & 0.09 & 0.07 & 0.08 & 0.52 & 0.52 & 0.52 \\
& $M=60$ & 0.14 & 0.10 & 0.12 & 0.52 & 0.52 & 0.52 \\
\midrule
\multirow{4}{*}{$4.0$}
& DP & 0.08 & 0.08 & 0.08 & 0.52 & 0.52 & 0.52 \\
& $M=40$ & 0.10 & 0.09 & 0.08 & 0.52 & 0.52 & 0.52 \\
& $M=50$ & 0.09 & 0.08 & 0.10 & 0.52 & 0.52 & 0.52 \\
& $M=60$ & 0.10 & 0.15 & 0.11 & 0.52 & 0.52 & 0.52 \\
\midrule
\multicolumn{8}{c}{\textbf{CIFAR-10}} \\
\midrule
Non-Private & -- 
& \multicolumn{3}{c}{1.14} 
& \multicolumn{3}{c}{0.62} \\
\midrule
\multirow{4}{*}{$2.0$} 
& DP & 0.09 & 0.09 & 0.09 & 0.51 & 0.51 & 0.51 \\
& $M=40$ & 0.13 & 0.12 & 0.10 & 0.51 & 0.51 & 0.51 \\
& $M=50$ & 0.13 & 0.12 & 0.08 & 0.51 & 0.51 & 0.51 \\
& $M=60$ & 0.13 & 0.14 & 0.13 & 0.51 & 0.51 & 0.51 \\
\midrule
\multirow{4}{*}{$4.0$} 
& DP & 0.08 & 0.08 & 0.08 & 0.50 & 0.50 & 0.50 \\
& $M=40$ & 0.07 & 0.10 & 0.08 & 0.50 & 0.50 & 0.50 \\
& $M=50$ & 0.09 & 0.07 & 0.09 & 0.50 & 0.50 & 0.51 \\
& $M=60$ & 0.11 & 0.09 & 0.14 & 0.50 & 0.50 & 0.50 \\
\bottomrule
\end{tabular}
\end{table}

\section{Handling GPU--TEE Discrepancy}
\label{section:gpu-tee-discrepancy}

\subsection{Honest Discrepancy and the Resulting Attack Surface}
Even when the GPU follows the protocol exactly, the clipped aggregate gradient it submits differs from the TEE's own recomputation on the same batch and the same trusted state. The two sides execute different kernels (vendor GPU libraries versus CPU BLAS), reduce sums in different orders, fuse operations differently, and round at different points; per-example clipping can then amplify a rounding-level difference in a single example's norm into a visible change of its clip factor. The discrepancy is therefore an intrinsic property of heterogeneous execution rather than a symptom of misbehavior, and a verifier that demanded bit-exact agreement would abort every honest run.

Two empirical properties of this discrepancy shape our design. Let $z_t^{32}=\Vert\bar g_t^{\mathrm{GPU}}-\bar g_t^{32}\Vert_2/C$ denote the normalized distance between the GPU submission and the TEE's FP32 recomputation. First, on the overwhelming majority of steps $z_t^{32}$ is tiny. Second, the distribution can have a heavy tail: rare, ill-conditioned steps---typically those containing examples whose per-example norm sits at the clipping boundary---produce discrepancies an order of magnitude larger than the body of the distribution. A single fixed tolerance is caught between two failure modes. Set near the body, it aborts honest runs on the tail steps; set near the tail, it hands a dishonest GPU a large per-step allowance on \emph{every} step.

Whatever tolerance the verifier grants for honest discrepancy is available to a dishonest GPU: a submission $\tilde g_t$ with $\Vert\tilde g_t-\bar g_t^{32}\Vert_2/C\le\tau$ is indistinguishable from an honest one on that step. Probabilistic checking already accounts for the fact that unchecked steps are not examined at all---this is the source of the $1-(1-p)^{M_{\mathrm{full}}}$ detection guarantee of Section~\ref{section:security_analysis}. Numerical tolerance adds a second, subtler channel: even a \emph{checked} step admits a deviation of up to $\tau$, and such deviations, being individually below the tolerance, would never be flagged. Left unmetered, they accumulate without bound over a long run, so the adversary could steer the model through many small, undetectable pushes rather than a few large ones. Our design principle is therefore that \emph{no accepted deviation is free}: every tolerated deviation observed on a checked step is charged to a bounded ledger, and hidden Bernoulli checking converts these sampled ledger charges into a high-probability bound on the cumulative deviation that can pass through the numerical-tolerance channels over the full run.

\subsection{Two-Stage Verification}
Rather than a single tolerance, the verifier uses an \emph{adaptive trusted reference}: TEE FP32 recomputation is the reference by default, and steps whose FP32 discrepancy is unusually large are escalated to a TEE FP64 recomputation. Every submission, checked or not, first passes a structural invariant: the honest average of clipped per-example gradients satisfies $\Vert\bar g_t^{\mathrm{GPU}}\Vert_2\le C$ by the triangle inequality, so the TEE rejects any submission with $\Vert\bar g_t^{\mathrm{GPU}}\Vert_2>C(1+10^{-4})$, where the slack covers only FP32 norm-computation rounding. This check costs $O(d)$, is applied on every step, and caps the gross deviation any single submission can carry. A checked step then follows one of three paths.

\smallskip\noindent\textit{Body path.} If $z_t^{32}\le\tau_{\mathrm{abs}}$, FP32 serves as the trusted reference. The verifier does \emph{not} simply accept: it charges the observed discrepancy to a cumulative ledger, $Z_{\mathrm{sub}}\leftarrow Z_{\mathrm{sub}}+z_t^{32}$, and aborts once $Z_{\mathrm{sub}}>K_{\mathrm{sub}}$. Thus $\tau_{\mathrm{abs}}$ is a routing threshold rather than a free tolerance: an adversary that repeatedly hides just below $\tau_{\mathrm{abs}}$ on checked steps exhausts $K_{\mathrm{sub}}$ after a bounded number of such steps, while honest runs, whose body discrepancies are far smaller than $\tau_{\mathrm{abs}}$, consume only a small fraction of the budget.

\smallskip\noindent\textit{Ambiguity path.} If $z_t^{32}>\tau_{\mathrm{abs}}$, the TEE recomputes the prescribed aggregate gradient in FP64, denoted $\bar g_t^{64}$, and evaluates $z_t^{64}=\Vert\bar g_t^{\mathrm{GPU}}-\bar g_t^{64}\Vert_2/C$, equivalently $n_t=(B/2)\,z_t^{64}$ in the accountant's units. The step is accepted only if $n_t\le\tau_{\mathrm{num}}$, i.e.\ $z_t^{64}\le\rho_{\mathrm{amb}}:=2\tau_{\mathrm{num}}/B$. The FP64 reference resolves the honest ill-conditioned cases, whose FP32 discrepancy was large only because FP32 rounding was amplified; but passing the FP64 check certifies plausibility, not honesty---an adversary may deliberately trigger escalation and hide within $\rho_{\mathrm{amb}}$. We therefore treat every accepted fallback as residual numerical ambiguity and meter it separately with a counter: $S_{\mathrm{amb}}\leftarrow S_{\mathrm{amb}}+1$, aborting once $S_{\mathrm{amb}}>K_{\mathrm{amb}}$. No assumption is made about why a step escalated or how many examples contributed to the discrepancy.

\smallskip\noindent\textit{Hard rejection.} If $z_t^{64}>\rho_{\mathrm{amb}}$, the submission is inconsistent with the prescribed computation under either reference and the verifier aborts.

\begin{algorithm}[t]
\small
\caption{TEE Verification}
\label{alg:gpu-tee-verification}

\KwIn{
GPU submission $\bar g_t^{\mathrm{GPU}}$;
verification probability $p$;
thresholds $\tau_{\mathrm{abs}},\tau_{\mathrm{num}}$;
limits $K_{\mathrm{sub}},K_{\mathrm{amb}}$;
clipping norm $C$;
denominator $B$.
}

GPU commits $\bar g_t^{\mathrm{GPU}}$\;

\If{$\Vert\bar g_t^{\mathrm{GPU}}\Vert_2>C(1+10^{-4})$}{
    \Return{\textsc{Abort}}\;
}

TEE secretly samples $V_t\sim\operatorname{Bernoulli}(p)$\;

\If{$V_t=0$}{
    \Return{\textsc{Accept}}\;
}

TEE recomputes $\bar g_t^{32}$ and sets
$z_t^{32}\leftarrow\Vert\bar g_t^{\mathrm{GPU}}-\bar g_t^{32}\Vert_2/C$\;

\If{$z_t^{32}\le\tau_{\mathrm{abs}}$}{
    $Z_{\mathrm{sub}}\leftarrow Z_{\mathrm{sub}}+z_t^{32}$\;
    \eIf{$Z_{\mathrm{sub}}>K_{\mathrm{sub}}$}{
        \Return{\textsc{Abort}}\;
    }{
        \Return{\textsc{Accept}}\;
    }
}

TEE recomputes $\bar g_t^{64}$ and sets
$n_t\leftarrow\frac{B}{2C}\Vert\bar g_t^{\mathrm{GPU}}-\bar g_t^{64}\Vert_2$\;

\If{$n_t>\tau_{\mathrm{num}}$}{
    \Return{\textsc{Abort}}\;
}

$S_{\mathrm{amb}}\leftarrow S_{\mathrm{amb}}+1$\;

\eIf{$S_{\mathrm{amb}}>K_{\mathrm{amb}}$}{
    \Return{\textsc{Abort}}\;
}{
    \Return{\textsc{Accept}}\;
}

\end{algorithm}

\smallskip
Algorithm~\ref{alg:gpu-tee-verification} summarizes the procedure. Two remarks are in order. First, deviation is always measured relative to the trusted reference the TEE actually produces, so the GPU's own numerical error never translates into attacker-controlled radius beyond what $\tau_{\mathrm{abs}}$ and $\rho_{\mathrm{amb}}$ explicitly grant; conversely, the honest spectrum is a property of the specific GPU--TEE software and hardware pair, and the parameters are frozen per calibrated pair. Second, for CIFAR-10 the honest FP32 spectrum has no heavy tail, so the FP64 layer is unnecessary: escalated steps are instead accepted only under the same-metric hard cap $z_t^{32}\le\rho_{\mathrm{ill}}=2\tau_{\mathrm{abs}}$ and counted by the same $K_{\mathrm{amb}}$ counter (Table~\ref{table:verifier-params}).

\subsection{Calibration and False Abort}
The four parameters play distinct roles: $\tau_{\mathrm{abs}}$ controls routing, $\rho_{\mathrm{amb}}$ controls FP64 acceptance, $K_{\mathrm{sub}}$ meters cumulative body discrepancy, and $K_{\mathrm{amb}}$ meters accepted fallback events. The verifier parameters are obtained through a staged calibration process. We first use a broad numerical-discrepancy campaign to characterize the body/tail structure of heterogeneous GPU--TEE execution and establish the two-stage calibration procedure. Before certified deployment, we apply this procedure to honest pilot trajectories on the target GPU--TEE pair, consolidate the resulting parameters at the model-family level, and freeze the complete verifier configuration for training. The verifier is calibrated for a concrete GPU--TEE numerical environment. Changes to the GPU family or trusted reference stack require recalibration; changes to the GPU-side training stack require revalidation and trigger recalibration if the resulting honest discrepancy spectrum falls outside the calibrated envelope. The verifier architecture, security accounting, and calibration procedure remain generic. The calibration trajectories are not identical in setup to the other experiments of this paper: they differ from the steering-attack experiments of Section~\ref{section:analysis_on_malicious_behavior} and the overhead runs of Section~\ref{section:experiment} in the weight-decay grouping (uniform versus the standard optimizer grouping; notes on Table~\ref{table:verifier-params}, Appendix~\ref{appendix:attack-adamw}), from the overhead runs additionally in the GPU-side software stack and the run length (Section~\ref{section:experiment}), and from the deployed protocol in the audit law under which the numerical data were recorded; Appendix~\ref{appendix:numerical-calibration} documents each difference and its effect. The honest false-abort probability $q_{\mathrm{FA}}$ is the probability, over the verifier's coins alone, that an honest execution of a given trajectory is aborted; it is a property of that trajectory, not a prediction for future runs. It is evaluated under the deployed Bernoulli verification law as a tail bound on the sampled body charge, a binomial term for escalations, and $(1-p)^{H_r}$ for hard rejections (Appendix~\ref{appendix:numerical-calibration}, Equation~\eqref{eq:false-abort-numerical}). For a fully observed honest trajectory it is evaluated \emph{conditionally on the realized numerical sequence} and requires no stationarity assumption; for trajectories observed only through a systematic every-tenth-step trace, the required trajectory statistics are estimated under an explicit representativeness assumption. Because $K_{\mathrm{sub}}$ and $K_{\mathrm{amb}}$ are absolute budgets while the honest audited charge grows with the number of steps, the reported values apply to runs of the calibrated length (Table~\ref{tab:attack-params}; $5{,}000$ steps for CIFAR-10); longer runs require rescaling $K_{\mathrm{sub}}$ and $K_{\mathrm{amb}}$, with the corresponding change in $G_{\mathrm{extra}}$. The design target is $q_{\mathrm{FA}}\le10^{-3}$ for every family. The family-level values in Table~\ref{table:verifier-params} are evaluated on the deployment-pair calibration trajectories themselves; they are therefore in-sample calibration checks, not estimates of the false-abort probability of any run. A false-abort estimate proper is available for one deployment trajectory only: after parameter freezing we preregistered a qqp-large trajectory with a fresh seed and replayed it with a full numerical census, obtaining an upper bound of $4\times10^{-100}$; no held-out trajectory exists for the GPT-2 and CIFAR-10 families, so no false-abort estimate is reported for them. A hypothetical $20\%$ inflation of the calibrated trajectory statistics keeps every family at or near this target, the thinnest margin being the GPT-2 body charge (Appendix~\ref{appendix:numerical-calibration}, which also gives the evidence tier of each reported quantity). The frozen verifier parameters, analytic security budgets, in-sample calibration checks, and the single held-out false-abort estimate are listed in Table~\ref{table:verifier-params}.

\subsection{Security Budget of Numerical Tolerance}
\label{sec:numerical-tolerance-security}

Numerical tolerance introduces a residual attack surface even on steps that would pass verification. Our ledgers meter such deviations whenever the
corresponding step is checked, while hidden Bernoulli sampling allows us to bound the total deviation that can pass through these tolerance channels over
the entire run.

Let $G_{\mathrm{tol}}$ denote the cumulative normalized deviation routed through the body and ambiguity channels. We derive a high-probability security budget
\begin{align}
G_{\mathrm{extra}}
=
G_{\mathrm{sub}}+G_{\mathrm{amb}},
\end{align}
such that
\begin{align}
\Pr[
G_{\mathrm{tol}}>G_{\mathrm{extra}}
\ \text{and still accepts}
]
\le
\beta_{\mathrm{extra}}
:=
\beta_{\mathrm{sub}}+\beta_{\mathrm{amb}}.
\label{eq:tolerance-security-pair}
\end{align}

For the normal FP32 path, $G_{\mathrm{sub}}$ follows from a Freedman--Bernstein bound for a predictable adversary that must commit before the current hidden Bernoulli coin is drawn. For the ambiguity path, $G_{\mathrm{amb}}$ follows from the exact binomial sampling law and the $K_{\mathrm{amb}}$ counter. Detailed derivations are given in Appendix~\ref{appendix:security_accounting}. At the primary operating point $p=0.1$, the resulting numerical-tolerance security pairs $(G_{\mathrm{extra}},\beta_{\mathrm{extra}})$ are $(5.48,0.05)$ for RoBERTa, $(9.50,0.05)$ for GPT-2, and $(2.51,0.05)$ for CIFAR-10 (Table~\ref{table:verifier-params}), where the first coordinate is the cumulative normalized-deviation budget and the second the corresponding security failure probability.

Importantly, Equation~\eqref{eq:tolerance-security-pair} is an analytic statement conditional on the frozen protocol parameters and the stated hidden-coin assumptions. Calibration determines the parameter values and hence the size of $G_{\mathrm{extra}}$; once they are frozen, the bound follows from the protocol and hidden Bernoulli sampling alone and does not assume that the adversarial discrepancy sequence follows the calibrated honest distribution. Appendix~\ref{appendix:steering-equivalence} further gives an empirical same-state steering-scale interpretation of this normalized budget; it is used only to interpret the effect of numerical tolerance on the full-step detection term (Section~\ref{section:security_analysis}), and the formal guarantee remains Equation~\eqref{eq:tolerance-security-pair}.

\section{Protocol Design}
\label{section:our_solution}

In this section, we design a verifiable training protocol that probabilistically detects repeated full deviations while accounting separately for deviations admitted through numerical tolerance.
The core idea of our design is \emph{probabilistic verification}. Instead of checking every training iteration, the TEE recomputes, or partially recomputes, gradients only on a small randomly selected subset of iterations. Since gradient recomputation inside the TEE is much slower than GPU execution, such checks must remain infrequent in order to keep the system efficient. However, the checking schedule is hidden from the GPU worker, so the adversary does not know in advance which iterations will be audited. Consequently, an adversary that deviates frequently faces a correspondingly high probability of being detected.

We now describe the basic protocol.

\textit{Initialization.} 
Before training begins, the TEE initializes and stores the trusted training state, including the model parameters, optimizer state, privacy parameters, and other global metadata. The GPU stores the training dataset and a local copy of the trainable parameters for efficient computation. Both parties share the public training specification, including the model architecture, loss function, optimizer, trainable parameter set, batch-selection procedure, and update rule.

\textit{Training.}
At the $t$-th iteration, the TEE sends the GPU the current trainable model state and the information needed to identify the current batch. The GPU fetches the batch, performs the forward and backward passes, and returns the gradients of the trainable parameters to the TEE. The TEE then uses hidden randomness to decide whether this iteration should be spot-checked.\footnote{The coin for step $t$ is drawn from TEE-internal entropy after the step-$t$ submission has been received. The measurement harness used for the experiments of Section~\ref{section:experiment} instead audited a hidden uniformly random subset of $\lceil pT\rceil$ steps per run, fixing each run's verification workload; Appendix~\ref{appendix:numerical-calibration} states what this affects.} If selected, the TEE recomputes, or partially recomputes, the expected gradient behavior on the same batch under the trusted state and compares it with the GPU output within an appropriate tolerance. 
If the discrepancy exceeds the tolerance, the TEE reports a deviation and terminates training. In addition, when persistent sealed storage and rollback protection are available, the TEE records a failure flag in its sealed state, so that subsequent attestations from the same TEE instance should not be certified.
If the check passes, the TEE continues with the trusted DP-SGD operations, including DP-noise generation, privacy/accounting-state maintenance, optimizer-state evolution, and parameter update. In the optimized implementation of Section~\ref{section:efficiency_optimization}, per-example clipping and aggregation are performed on the GPU and probabilistically verified by trusted recomputation. The updated trainable state is then sent back to the GPU for the next iteration. 

\textit{Output.}
After training completes and every queued verification job has finished, the TEE releases the final model and attaches a TEE-generated certificate indicating that all audited iterations were consistent with the declared training protocol and that the numerical ledgers remained within their limits.
The overall protocols are summarized in Algorithm \ref{alg:verifiable_dpsgd_gpu} and \ref{alg:verifiable_dpsgd_tee}.

Overall, our design separates efficiency-critical computation from
trust-critical computation. The GPU performs the expensive training-side
operations, while the TEE retains control over verification and privacy
enforcement. In this way, the protocol substantially reduces the TEE-side
burden compared with fully trusted training, while making repeated full
deviations risky and explicitly accounting for manipulation admitted through
numerical tolerance. In the following sections, we analyze the security guarantees of the protocol and introduce several optimizations that further reduce verification overhead without compromising security.

\begin{algorithm}[!t]
\small
\caption{GPU-side Gradient Computation and Local Update}
\label{alg:verifiable_dpsgd_gpu}

\KwIn{
Training dataset $\mathcal{D}$;
model $f(\cdot;\boldsymbol{\theta})$ and loss $\ell$;
trainable parameters $\mathcal{T}$;
clipping norm $C$;
optimizer rule $\mathsf{OptStep}$;
batch seed $s_{\mathrm{batch}}$;
initial weights $\boldsymbol{\theta}^{(0)}_{\mathcal{T}}$;
initial optimizer state $\boldsymbol{\omega}^{(0)}$;
number of iterations $T$;
aggregation denominator $B$.
}

\textbf{GPU:} store $\mathcal{D}$ and receive
$(\boldsymbol{\theta}^{(0)}_{\mathcal{T}},
\boldsymbol{\omega}^{(0)},s_{\mathrm{batch}})$ from the TEE\;

\textbf{GPU:} set
$\widehat{\boldsymbol{\theta}}^{(0)}_{\mathcal{T}}
\leftarrow\boldsymbol{\theta}^{(0)}_{\mathcal{T}}$
and
$\widehat{\boldsymbol{\omega}}^{(0)}
\leftarrow\boldsymbol{\omega}^{(0)}$\;

\BlankLine

\For{$t\leftarrow0$ \KwTo $T-1$}{

    $I_t\leftarrow\mathsf{BatchIdx}(s_{\mathrm{batch}},t)$\;
    \textbf{GPU:} load $B_t\leftarrow\mathcal{D}[I_t]$\;

    \ForEach{$(x_i,y_i)\in B_t$}{
        $\boldsymbol{g}_{t,i}\leftarrow
        \nabla_{\boldsymbol{\theta}_{\mathcal{T}}}
        \ell(f(x_i;\widehat{\boldsymbol{\theta}}^{(t)}),y_i)$\;

        $\bar{\boldsymbol{g}}_{t,i}\leftarrow
        \boldsymbol{g}_{t,i}
        \min\{1,C/\|\boldsymbol{g}_{t,i}\|_2\}$\;
    }

    $\bar{\boldsymbol{g}}_t
    \leftarrow
    \frac{1}{B}
    \sum_{(x_i,y_i)\in B_t}
    \bar{\boldsymbol{g}}_{t,i}$\;

    \textbf{GPU $\rightarrow$ TEE:} commit and send
    $(t,I_t,\bar{\boldsymbol{g}}_t)$\;

    \textbf{GPU:} wait for the TEE to release $s_{\mathrm{noise},t}$\;

    $\widetilde{\boldsymbol{g}}_t
    \leftarrow
    \bar{\boldsymbol{g}}_t+\frac{1}{B}
    \mathsf{Noise}(s_{\mathrm{noise},t};\sigma C)$\;

    $(\widehat{\boldsymbol{\theta}}^{(t+1)}_{\mathcal{T}},
    \widehat{\boldsymbol{\omega}}^{(t+1)})
    \leftarrow
    \mathsf{OptStep}\!\left(
    \widehat{\boldsymbol{\theta}}^{(t)}_{\mathcal{T}},
    \widehat{\boldsymbol{\omega}}^{(t)},
    \widetilde{\boldsymbol{g}}_t
    \right)$\;
}

\end{algorithm}

\begin{algorithm}[!t]
\small
\caption{TEE-side Verification, Noise Release, and Update}
\label{alg:verifiable_dpsgd_tee}

\KwIn{
Training dataset $\mathcal{D}$;
model $f(\cdot;\boldsymbol{\theta})$ and loss $\ell$;
trainable parameters $\mathcal{T}$;
DP parameters $(C,\sigma)$;
optimizer rule $\mathsf{OptStep}$;
verification probability $p$;
number of iterations $T$;
aggregation denominator $B$.
}

\KwOut{
Final trusted weights
$\boldsymbol{\theta}^{(T)}_{\mathcal{T}}$.
}

\textbf{TEE:} initialize trusted weights
$\boldsymbol{\theta}^{(0)}_{\mathcal{T}}$
and optimizer state $\boldsymbol{\omega}^{(0)}$\;

\textbf{TEE:} choose master seeds
$s_{\mathrm{batch}}$ and $s_{\mathrm{noise}}$\;

\textbf{TEE $\rightarrow$ GPU:} send
$(\boldsymbol{\theta}^{(0)}_{\mathcal{T}},
\boldsymbol{\omega}^{(0)},
s_{\mathrm{batch}})$\;

\BlankLine

\For{$t\leftarrow0$ \KwTo $T-1$}{

    \textbf{TEE:} receive committed
    $(t,I_t,\bar{\boldsymbol{g}}^{\mathrm{GPU}}_t)$ from the GPU\;

    \textbf{TEE:} check
    $I_t=\mathsf{BatchIdx}(s_{\mathrm{batch}},t)$\;

    draw $V_t\sim\operatorname{Bernoulli}(p)$ secretly\;

    \If{$V_t=1$}{
        \textbf{TEE:} load $B_t\leftarrow\mathcal{D}[I_t]$\;

        \ForEach{$(x_i,y_i)\in B_t$}{
            $\boldsymbol{g}^{\mathrm{TEE}}_{t,i}
            \leftarrow
            \nabla_{\boldsymbol{\theta}_{\mathcal{T}}}
            \ell(f(x_i;\boldsymbol{\theta}^{(t)}),y_i)$\;

            $\bar{\boldsymbol{g}}^{\mathrm{TEE}}_{t,i}
            \leftarrow
            \boldsymbol{g}^{\mathrm{TEE}}_{t,i}
            \min\{1,C/\|\boldsymbol{g}^{\mathrm{TEE}}_{t,i}\|_2\}$\;
        }

        $\bar{\boldsymbol{g}}^{\mathrm{TEE}}_t
        \leftarrow
        \frac{1}{B}
        \sum_{(x_i,y_i)\in B_t}
        \bar{\boldsymbol{g}}^{\mathrm{TEE}}_{t,i}$\;

        \textbf{TEE:} perform numerical verification using $\bar{\boldsymbol{g}}^{\mathrm{TEE}}_t$\;

        \If{verification aborts}{
            \Return{\textsc{Abort}}\;
        }
    }

    $s_{\mathrm{noise},t}
    \leftarrow
    \mathsf{Derive}(s_{\mathrm{noise}},t)$\;

    \textbf{TEE $\rightarrow$ GPU:} release $s_{\mathrm{noise},t}$\;

    $\boldsymbol{\xi}_t
    \leftarrow
    \frac{1}{B}
    \mathsf{Noise}(s_{\mathrm{noise},t};\sigma C)$\;

    $\widetilde{\boldsymbol{g}}_t
    \leftarrow
    \bar{\boldsymbol{g}}^{\mathrm{GPU}}_t
    +
    \boldsymbol{\xi}_t$\;

    $(\boldsymbol{\theta}^{(t+1)}_{\mathcal{T}},
    \boldsymbol{\omega}^{(t+1)})
    \leftarrow
    \mathsf{OptStep}\!\left(
    \boldsymbol{\theta}^{(t)}_{\mathcal{T}},
    \boldsymbol{\omega}^{(t)},
    \widetilde{\boldsymbol{g}}_t
    \right)$\;
}

\Return{$\boldsymbol{\theta}^{(T)}_{\mathcal{T}}$}\;

\end{algorithm}

Algorithm~\ref{alg:verifiable_dpsgd_tee} is the blocking reference execution. The deployed hungry-updating implementation keeps the same certificate-time accept/reject semantics but reorders the runtime: once the submission is committed and the coin is sampled, a selected check is enqueued for background recomputation while the TEE proceeds with the provisional trusted update and seed release; the state remains provisional until every deferred job has completed successfully (Section~\ref{section:efficiency_optimization}).

\begin{figure*}
\centering
\includegraphics[width=1.0\textwidth]{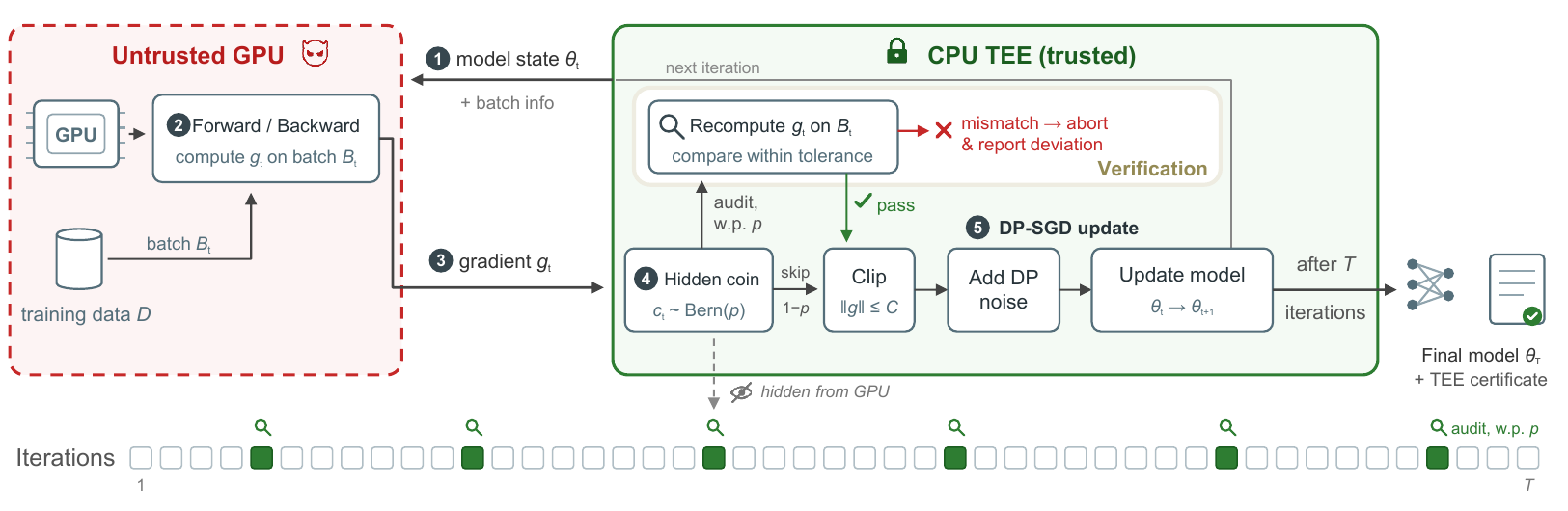}
\caption{Overview of our basic verifiable DP training protocol. The untrusted GPU performs the expensive forward and backward computation (steps 1--3), while the TEE uses a hidden coin $V_t \sim \mathrm{Bernoulli}(p)$ to decide whether to audit the returned gradient by recomputation (step 4), and then executes the privacy-critical DP-SGD update (step 5). After $T$ iterations, the TEE releases the final model with an attestation certificate. The figure shows the logical baseline; in the deployed implementation (Section~\ref{section:efficiency_optimization}, Algorithms~\ref{alg:verifiable_dpsgd_gpu}--\ref{alg:verifiable_dpsgd_tee}) clipping and aggregation run on the GPU and are recomputed by the TEE only on audited steps. Since each iteration is audited independently with probability $p$ and the schedule is hidden from the GPU, repeated full deviations are detected with increasing probability while the GPU retains its efficiency advantage.}
\label{fig:basic protocol}
\end{figure*}

\subsection{Security Analysis \& Incentive Engineering}
\label{section:security_analysis}

\paragraph{Accounting for numerical tolerance.}
An adversary may combine full deviations with deviations that remain inside
the numerical-tolerance region; we account for the two components
separately, each established on its own terms. First,
deviations that a check would reject---submissions outside the tolerated
region on their step---are \emph{full deviations}. For a fixed set $\mathcal F$ of them,
\begin{align}
P_{\mathrm{det}}(\mathcal F)
=
1-
\prod_{t\in\mathcal F}(1-p_t).
\label{eq:full-detection}
\end{align}
Under the uniform policy $p_t=p$ used throughout this work, the same argument applies to the first $M_{\mathrm{full}}$ predictable full-deviation opportunities generated by a history-adaptive adversary: each opportunity is determined before its hidden verification coin is drawn,\footnote{The coin is drawn only after the submission arrives, so no current-step decision exists at commitment time. What the GPU can observe is whether the \emph{previous} step triggered extra TEE work (a timing channel); under independent coins this reveals nothing about later steps. Our analysis abstracts away timing and other side channels that could expose the current decision before commitment; a deployment can harden this by padding the latency-critical response path or by deferring verification dispatch to an asynchronous queue after commitment.} so all $M_{\mathrm{full}}$ opportunities are missed with probability $(1-p)^{M_{\mathrm{full}}}$ and $P_{\mathrm{det}}=1-(1-p)^{M_{\mathrm{full}}}$. Second,
deviations \emph{within} the tolerated region are never individually
flagged, but Section~\ref{sec:numerical-tolerance-security} bounds their
total: the cumulative normalized deviation accepted through the tolerance
channels over an entire run exceeds $G_{\mathrm{extra}}$ without the run
being aborted with probability at most
$\beta_{\mathrm{extra}}=\beta_{\mathrm{sub}}+\beta_{\mathrm{amb}}$.
The full-deviation bound and the pair
$(G_{\mathrm{extra}},\beta_{\mathrm{extra}})$ together constitute the formal
security guarantee. They are stated in different units, require no
empirical conversion between them, and both forms of deviation may coexist
in the same execution. The uniform-$p$ result and the pair hold for history-adaptive strategies satisfying the
commit-before-current-coin condition; the nonuniform value-aware extension of Appendix~\ref{appendix:value-aware} is stated for a fixed set of full deviations.

\paragraph{Scope of the guarantee.}
Every committed submission that passes the structural check falls,
according to the verdict a check would return, into exactly one of three
classes: full (rejected-class), sub-threshold body, or ambiguity-path
deviation. The uniform-$p$ full-deviation bound and $(G_{\mathrm{extra}},\beta_{\mathrm{extra}})$
therefore account for arbitrary submitted gradients in the verifier's
normalized-deviation metric, including strategies that adapt their direction,
magnitude, or timing to the previous protocol history, subject to the
commit-before-current-coin condition.

What the formal accounting does not determine is how a given normalized
deviation translates into optimization progress or model utility. Such an
effect may depend strongly on the optimizer state, attack direction, and
training phase. Appendix~\ref{appendix:steering-equivalence} provides only an
empirical calibration of this conversion over the tested attack states and
radii. Attacks through channels other than the submitted aggregate gradient
remain outside the present analysis.

The formal quantities above do not depend on the attack experiments.
The empirical operating point $\overline{M}=50$
(Section~\ref{section:analysis_on_malicious_behavior}) and the steering-scale
conversion of Appendix~\ref{appendix:steering-equivalence} are measured on
the attack-experiment trajectories, whose setup differs from that of the
calibration trajectories
(Section~\ref{section:gpu-tee-discrepancy}). Interpreting these empirical
findings together with the security budgets instantiated for the deployment
pair therefore assumes that their qualitative conclusions transfer across
these trajectory families. We treat this as an empirical transfer assumption,
not as part of the formal security guarantee. More generally, whenever an
empirical quantity measured on one trajectory family is used elsewhere in the
paper to interpret results instantiated on another trajectory family, the same
transfer assumption should be understood unless stated otherwise.

\paragraph{Empirical step-equivalent interpretation.}
Appendix~\ref{appendix:steering-equivalence} calibrates one normalized deviation unit against one full-power steering update and finds a ratio close to one over the tested regime. Measure each accepted deviation by its norm in units of the clipping bound: a step on which the honest aggregate is replaced by a full-power forgery contributes one unit, and a tolerated deviation of normalized radius $r$ contributes $r$ units, which is what the calibration above establishes. Writing $A_{\mathrm{steer}}$ for the sum of these contributions over a run --- an attack's full-power-equivalent steering magnitude, real-valued and distinct from the step count $M_{\mathrm{full}}$, though an attack that forges at full power on $M_{\mathrm{full}}$ steps and nowhere else has $A_{\mathrm{steer}}=M_{\mathrm{full}}$ --- the tolerance channels contribute at most about $G_{\mathrm{extra}}$ units, leaving $(A_{\mathrm{steer}}-G_{\mathrm{extra}})_{+}$ units outside the budget:
\begin{align}
P^{\mathrm{interp}}_{\mathrm{det}}(A_{\mathrm{steer}})
\approx
1-(1-p)^{(A_{\mathrm{steer}}-G_{\mathrm{extra}})_{+}}.
\label{eq:effective-detection}
\end{align}
Equation~\eqref{eq:effective-detection} is an interpretation: it rests on the empirical calibration and on treating tolerated and full deviations additively, it does not include $\beta_{\mathrm{extra}}$, which is reported separately, and it interprets the cost of numerical tolerance only for attacks whose steering behaviour is represented by the calibration of Appendix~\ref{appendix:steering-equivalence}, not as a worst-case guarantee over arbitrary attack objectives. With $G_{\mathrm{extra}}=0$ and $A_{\mathrm{steer}}=M_{\mathrm{full}}$ it reduces to $1-(1-p)^{M_{\mathrm{full}}}$. For illustration, with $A_{\mathrm{steer}}=40$, $p=0.1$, and the conservative $G_{\mathrm{extra}}=10$, the term decreases from $98.52\%$ to $95.76\%$; under our calibrated budgets ($G_{\mathrm{extra}}\le9.5$) the change is smaller.

Attack value may vary across training steps, but this timing information is also available to the verifier, which can allocate more of the same expected checking budget to the more sensitive phases so that higher-value attack steps also carry higher detection risk (Appendix~\ref{appendix:value-aware}). All budgets and experiments in this paper use the uniform setting $p_t=p$.

\subsection{Efficiency Optimizations}
\label{section:efficiency_optimization}
To make the split TEE-GPU design practical, we optimize both the communication path and the execution pipeline, which can substantially reduce the end-to-end overhead of verifiable DP training.

\vspace{2pt}
\noindent\textbf{Communication-efficient split execution.}
A straightforward implementation of the basic protocol would send all per-example gradients to the TEE and perform clipping inside the trusted environment. For a batch of size $B$ and model dimension $d$, this requires $O(Bd)$ GPU-to-TEE communication per iteration, which can dominate the runtime for modern models.

We instead perform clipping on the GPU. The GPU computes the per-example gradients, clips each gradient, averages and submits only the resulting clipped-and-averaged gradient $\bar g_t^{\mathrm{GPU}}$ to the TEE. The TEE retains responsibility for the privacy-critical randomness and optimizer state, while probabilistic verification checks whether the GPU-submitted aggregate is consistent with the prescribed computation. This reduces the GPU-to-TEE communication from $O(Bd)$ to $O(d)$ per iteration.

We also avoid sending the updated model and optimizer state back to the GPU after every step. After the GPU commits $\bar g_t^{\mathrm{GPU}}$, the TEE releases the random seed used to generate the trusted DP noise for that step. The GPU reconstructs the same noise locally and applies the same optimizer update, thereby maintaining a synchronized copy of the model and optimizer state. The TEE-to-GPU communication is therefore reduced from $O(d)$ to the size of a random seed. Releasing the seed only after the gradient commitment is essential: the GPU cannot adapt its submitted gradient to the realized DP noise.

Thus, the common training path exchanges only one aggregate gradient in the GPU-to-TEE direction and a short random seed in the reverse direction. The more expensive per-example recomputation is incurred only on the subset of iterations selected for verification.

\vspace{2pt}
\noindent\textbf{Hungry updating with deferred verification.}
Gradient verification is substantially more expensive than the ordinary trusted update. In particular, a checked iteration requires the TEE to recompute the prescribed aggregate gradient in FP32, and a step exhibiting a large FP32 discrepancy may additionally require FP64 adjudication. Performing these computations synchronously would stall the GPU whenever a verification is triggered and would largely eliminate the benefit of probabilistic checking.

We therefore decouple model updating from gradient verification using \emph{hungry updating}. Once the GPU commits the aggregate gradient, the TEE executes the latency-critical DP update without waiting for the corresponding gradient-honesty check to finish. If the iteration is selected for verification, the TEE records the state needed to reproduce that iteration and places a verification job into a background queue. Training can then continue while background workers independently recompute the checked iterations. Because verification is deferred, a checked submission may enter the provisional training state before its job completes; such progress is not certified. A run is \emph{accepted}---the meaning of the term throughout the security analysis---only when every verification job generated during the run has completed successfully and the certificate is issued; any failed deferred check aborts the run.

Each queued job executes the verification procedure described in Section~\ref{section:gpu-tee-discrepancy}.  Multiple verification workers can serve the queue in parallel, allowing training to run ahead of verification when sufficient CPU resources are available.

\begin{figure}
\centering
\includegraphics[width=\columnwidth]{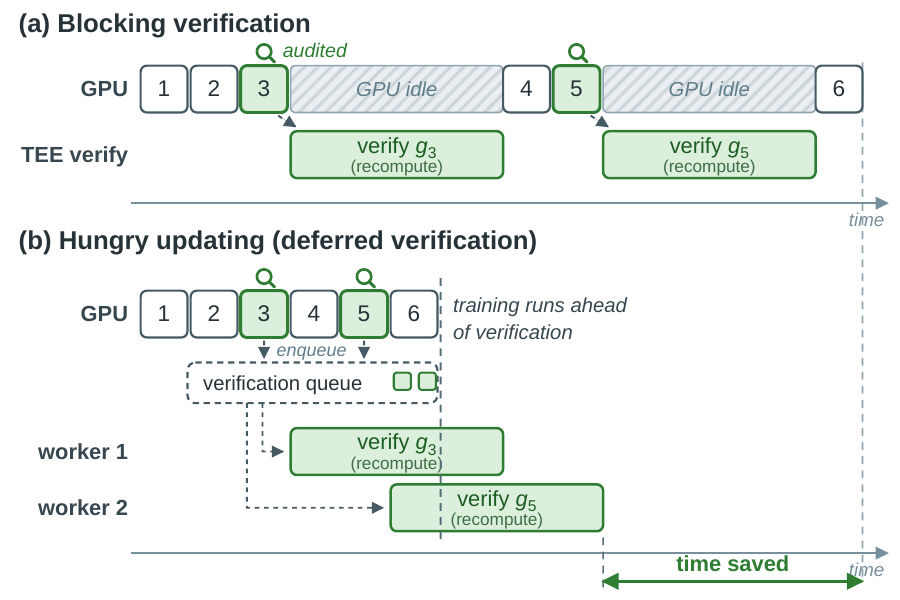}
\caption{
Hungry updating with deferred verification.
(a) Blocking verification stalls training whenever a checked iteration is recomputed by the TEE.
(b) Hungry updating places checked iterations into a background queue served by parallel verification workers, allowing training to proceed while verification executes asynchronously.
}
\label{fig:hungry_update}
\end{figure}

Figure~\ref{fig:hungry_update} illustrates the resulting pipeline. With these optimizations, the normal training path incurs only lightweight communication and trusted updating, while expensive recomputation is parallelized over probabilistically selected verification steps. In particular,
when the aggregate verification capacity keeps pace with the arrival of checking tasks, i.e., $n \ge\frac{pT_{\mathrm{iter\text{-}veri}}}{T_{\mathrm{iter}}}$,
most of the verification overhead can be hidden behind the main training pipeline, leaving only a small queue-draining cost at the end of training. Otherwise, verification becomes the throughput bottleneck. We provide a detailed workload and efficiency analysis in Appendix~\ref{appendix:efficiency_analysis}.

\vspace{-0.5cm}

\section{Experiments}
\label{section:experiment}

We conduct experiments on multiple datasets and models to evaluate our system. 
The evaluated models and datasets follow Section~\ref{section:analysis_on_malicious_behavior} and the details are provided in Appendix~\ref{appendix:data_description}.

\subsection{Experimental setup}

\begin{table*}[t] \centering \small \setlength{\tabcolsep}{4pt} \caption{ End-to-end overhead and runtime diagnostics at two verification rates, excluding initialization and prewarming. $T_{\mathrm{In\text{-}GPU}}$ is the one-epoch unverified in-GPU DP training time. $T_{\mathrm{VDP}}$ is the one-epoch training time of our protocol. $T_{\mathrm{In\text{-}TEE}}$ represents the estimated one-epoch training time of the full in-TEE solution. CIFAR-10 rows use a 200-step window for $T_{\mathrm{VDP}}$ against a 500-step In-GPU baseline (its checks are $\sim$20$\times$ costlier, hence the lower rates $p\in\{0.02,0.05\}$); the VDP overhead is therefore computed per warm step, excluding the first compiled step on both sides (per warm step: In-GPU $2.515$\,s versus VDP $2.843$\,s at $p{=}0.02$ and $3.284$\,s at $p{=}0.05$; the CIFAR-10 In-TEE ratio is also taken per warm step, and the raw totals in the CIFAR-10 row are not directly comparable across columns). For the LLM operating rates $p{=}0.10$ and $p{=}0.15$ shown here, $M_{\mathrm{full}}=50$ full deviations are detected with probability $1-0.9^{50}=99.5\%$ and $1-0.85^{50}=99.97\%$, respectively (Eq.~\eqref{eq:full-detection}); the CIFAR-10 rows use the lower rates $p\in\{0.02,0.05\}$ and are to be read with Eq.~\eqref{eq:full-detection} at those rates; in addition, for the LLM cells the cumulative deviation admitted through the numerical-tolerance channels is bounded by $(G_{\mathrm{extra}},\beta_{\mathrm{extra}})=(9.5,0.05)$ or better (Section~\ref{section:security_analysis}; the budgets of Table~\ref{table:verifier-params} are computed at the primary operating point $p{=}0.1$ and are conservative at $p{=}0.15$).}\label{tab:overhead-breakdown} \begin{tabular}{lllrrrrrrrr} \toprule Model & Task & $p$ & $T_{\mathrm{In\text{-}GPU}}$ (s) & $T_{\mathrm{VDP}}$ (s) & $T_{\mathrm{In\text{-}TEE}}$ (s) & Sync (s) & Drain (s) & Backpr. (s) & VDP Overhead & In-TEE\\ \midrule \multirow{2}{*}{WRN16-4} & \multirow{2}{*}{CIFAR-10} & 0.02 & \multirow{2}{*}{1306.04} & 743.29 & \multirow{2}{*}{25160.00} & 203.98 & 0.00 & 0.00 & $+13.0\%$ & \multirow{2}{*}{$20.0\times$} \\ & & 0.05 & & 801.80 & & 234.23 & 5.47 & 509.44 & $+30.6\%$ & \\ \midrule \multirow{4}{*}{GPT-2 Small} & \multirow{2}{*}{WebNLG} & 0.10 & \multirow{2}{*}{109.12} & 115.24 & \multirow{2}{*}{1001.03} & 5.11 & 0.00 & 0.40 & $+5.6\%$ & \multirow{2}{*}{$9.2\times$} \\ & & 0.15 & & 122.01 & & 12.04 & 2.28 & 7.08 & $+11.8\%$ & \\ & \multirow{2}{*}{E2E} & 0.10 & \multirow{2}{*}{119.08} & 127.29 & \multirow{2}{*}{1095.97} & 6.32 & 3.13 & 0.93 & $+6.9\%$ & \multirow{2}{*}{$9.2\times$} \\ & & 0.15 & & 140.20 & & 15.32 & 4.01 & 10.23 & $+17.7\%$ & \\ \midrule \multirow{4}{*}{GPT-2 Medium} & \multirow{2}{*}{WebNLG} & 0.10 & \multirow{2}{*}{277.84} & 290.53 & \multirow{2}{*}{2298.14} & 8.21 & 0.00 & 0.71 & $+4.6\%$ & \multirow{2}{*}{$8.3\times$} \\ & & 0.15 & & 310.39 & & 27.26 & 5.75 & 20.80 & $+11.7\%$ & \\ & \multirow{2}{*}{E2E} & 0.10 & \multirow{2}{*}{301.79} & 315.11 & \multirow{2}{*}{2548.66} & 10.52 & 5.85 & 1.50 & $+4.4\%$ & \multirow{2}{*}{$8.4\times$} \\ & & 0.15 & & 344.75 & & 31.48 & 10.14 & 22.52 & $+14.2\%$ & \\ \midrule \multirow{4}{*}{RoBERTa Base} & \multirow{2}{*}{QQP} & 0.10 & \multirow{2}{*}{582.96} & 667.27 & \multirow{2}{*}{5003.06} & 47.09 & 0.36 & 7.97 & $+14.5\%$ & \multirow{2}{*}{$8.6\times$} \\ & & 0.15 & & 723.43 & & 93.34 & 0.74 & 48.00 & $+24.1\%$ & \\ & \multirow{2}{*}{MNLI} & 0.10 & \multirow{2}{*}{631.72} & 726.14 & \multirow{2}{*}{5022.67} & 53.34 & 0.84 & 20.56 & $+14.9\%$ & \multirow{2}{*}{$8.0\times$} \\ & & 0.15 & & 797.76 & & 117.26 & 1.47 & 86.86 & $+26.3\%$ & \\ \midrule \multirow{4}{*}{RoBERTa Large} & \multirow{2}{*}{QQP} & 0.10 & \multirow{2}{*}{1724.31} & 1911.31 & \multirow{2}{*}{15590.00} & 109.57 & 0.97 & 61.53 & $+10.8\%$ & \multirow{2}{*}{$9.0\times$} \\ & & 0.15 & & 2176.47 & & 361.03 & 8.97 & 344.72 & $+26.2\%$ & \\ & \multirow{2}{*}{MNLI} & 0.10 & \multirow{2}{*}{1873.26} & 2125.79 & \multirow{2}{*}{15863.42} & 173.04 & 4.28 & 136.99 & $+13.5\%$ & \multirow{2}{*}{$8.5\times$} \\ & & 0.15 & & 2439.39 & & 477.86 & 10.58 & 481.74 & $+30.2\%$ & \\ \bottomrule \end{tabular} \end{table*}

\noindent \textbf{Hardware.}
We instantiate the untrusted trainer with a single NVIDIA RTX PRO 6000 GPU. The TEE-side verifier runs in an AMD SEV-SNP-protected VM configured with 64 CPUs and 512GB of memory. The GPU and the TEE communicate over a vsock channel~\cite{russell2008virtio}.

\smallskip \noindent \textbf{Verification workflow.}
Checks are dispatched asynchronously so that verification overlaps GPU training. The judge parameters are frozen per model family --- one configuration for all GPT-2 tasks, one for all RoBERTa tasks, and one for CIFAR-10 (Table~\ref{table:verifier-params}). The resulting adversarial budgets at the primary operating point $p{=}0.1$ are $G_{\mathrm{extra}}\le5.48$ (RoBERTa), $9.50$ (GPT-2), and $2.51$ (CIFAR-10), each with $\beta_{\mathrm{extra}}=0.05$. The false-abort probability is estimated on one held-out deployment trajectory only, the preregistered qqp-large full-census replay run after the parameters were frozen, which gives an upper bound of $4\times10^{-100}$ against the design target $10^{-3}$. The family-level tail values on the deployment-pair calibration trajectories, $2.8\times10^{-9}$ (RoBERTa), $8.7\times10^{-6}$ (GPT-2), and $1.1\times10^{-4}$ (CIFAR-10) under their respective evidence models (Appendix~\ref{appendix:numerical-calibration}), confirm that the frozen parameters meet the target on seven of the eight calibration trajectories (the qqp-large calibration trace is excluded, see Appendix~\ref{appendix:numerical-calibration}); they are in-sample calibration checks, not false-abort estimates. Generalization to future trajectories remains empirical. The overhead runs reported here differ from the deployment-pair calibration trajectories in three respects: the GPU-side software stack (the deployed PyTorch and PEFT versions, whereas the RoBERTa-base and GPT-2 calibration trajectories were produced under an earlier build), the optimizer grouping (the standard HF grouping that exempts biases and LayerNorm weights from weight decay, whereas the calibration trajectories apply it uniformly; notes on Table~\ref{table:verifier-params}), and the run length (one epoch under the ten-epoch schedule rather than full trajectories). Appendix~\ref{appendix:numerical-calibration} details the software-stack difference and revalidates the frozen configuration against the honest discrepancy spectrum of these runs. The additional verification rates in Table~\ref{tab:overhead-breakdown} characterize the efficiency--rate tradeoff only; the budgets are not transferred to other rates without recomputation (with the same counters, the CIFAR-10 pair becomes $(5.05,0.05)$ at $p{=}0.05$ and $(12.66,0.05)$ at $p{=}0.02$).

\smallskip \noindent \textbf{Benchmarks and baselines.}
We measure system overhead on both traditional learning tasks (CIFAR-10) and LLM-oriented tasks (E2E, WebNLG, MNLI, QQP), using the same family of models as in Section~\ref{section:analysis_on_malicious_behavior}. We omit Purchase because its training schedule is too short to yield meaningful epoch-level timing. Our floor-level baseline is unverified DP training on a single GPU, which incurs no communication or checking cost. We also report a \emph{TEE-only} baseline that runs the entire DP training procedure inside the CPU-TEE, representing the naive alternative. It is implemented as a single-machine CPU DP-SGD trainer inside the same SEV-SNP guest, reusing the exact DP recipe of the verified runs (per-example clipping, seeded noise, optimizer, and schedule) with the batch sharded across the guest's vCPUs at the measured optimal shape (32 workers $\times$ 2 threads). Because a full in-TEE epoch takes hours to days, we time five optimizer steps after a warm-up step and extrapolate linearly to one epoch. For CIFAR-10 the in-TEE time is derived from the measured in-guest cost of the verifier's full-batch recomputation (the same computation as one training step) on all 64 vCPUs. The measurement harness audits exactly $\lceil pT\rceil$ steps per run, i.e., the protocol's mean checking workload; the reported times therefore exclude the run-to-run variation that the Bernoulli check count (relative standard deviation $\sqrt{(1-p)/(pT)}$, $6$--$13\%$ on the LLM cells and $31$--$49\%$ on the CIFAR-10 windows) and the placement of checked steps would add.

\smallskip
\noindent\textbf{Metrics.}
We report wall-clock timing for a single training epoch, excluding initialization and worker prewarming. $T_{\mathrm{In\text{-}GPU}}$ denotes the runtime of unverified in-GPU DP training on the same GPU, while $T_{\mathrm{VDP}}$ denotes the end-to-end runtime of our protocol, including the final completion of all verification jobs generated during the epoch. We report the following diagnostics:
\begin{itemize}
    \item \textit{Sync}, the cumulative GPU-side waiting time after submitting an update and before receiving the corresponding noise seed from the TEE. It includes gradient serialization, GPU--TEE communication, deserialization, and TEE response latency, and is therefore an upper bound on pure communication cost.

    \item \textit{Drain}, the time required to complete verification jobs that remain pending after GPU training finishes. This captures the verification work that cannot be hidden behind training.

    \item \textit{Backpr.}, the time training is blocked because the number of in-flight verification jobs reaches the protocol limit. A nonzero value indicates that verification throughput becomes a bottleneck. Since Backpr. is measured on the TEE side while Sync is measured on the GPU side, the two may overlap and should not be added directly.

    \item \textit{End-to-End Overhead}, the end-to-end slowdown relative to unverified in-GPU training.
\end{itemize}

\subsection{Efficiency Evaluation}

\begin{table}[!b]
\centering
\caption{Model accuracy of WideResNet on CIFAR-10 under different privacy budgets with $M=120$ malicious iterations.}
\label{tab:acc_cifar10_30_60_120}
\small
\begin{tabular}{ccccc}
\toprule
\small
$\varepsilon$ & Setting & Current & Nearest & Final \\
\midrule
$\infty$ & Non-Private & \multicolumn{3}{c}{92.38\%} \\
\midrule
\multirow{2}{*}{$2.0$}
& DP & 53.48\% & 53.48\% & 53.48\% \\
& $M=120$ & 54.20\% & 52.81\% & 53.83\% \\
\midrule

\multirow{2}{*}{$4.0$}
& DP & 65.80\% & 65.80\% & 65.80\% \\
& $M=120$ & 65.09\% & 64.83\% & 65.81\% \\
\midrule

\multirow{2}{*}{$8.0$}
& DP & 72.23\% & 72.23\% & 72.23\% \\
& $M=120$ & 83.21\% & 72.99\% & 85.38\% \\
\bottomrule
\end{tabular}
\end{table}

\begin{figure}[t]
  \centering
  \begin{subfigure}[b]{\columnwidth}
    \centering
    \scalebox{0.8}[0.65]{%
      \includegraphics[width=\columnwidth]{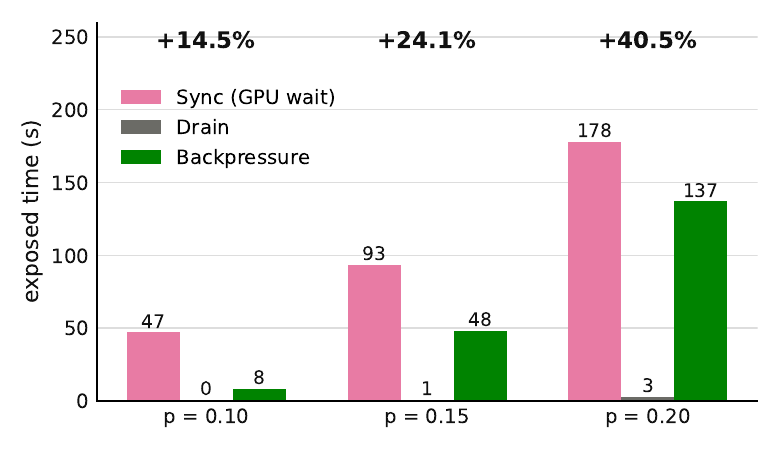}}
    \caption{QQP / RoBERTa-base}
    \label{fig:regime-base}
  \end{subfigure}
  \vspace{2pt}
  \begin{subfigure}[b]{\columnwidth}
    \centering
    \scalebox{0.8}[0.65]{%
      \includegraphics[width=\columnwidth]{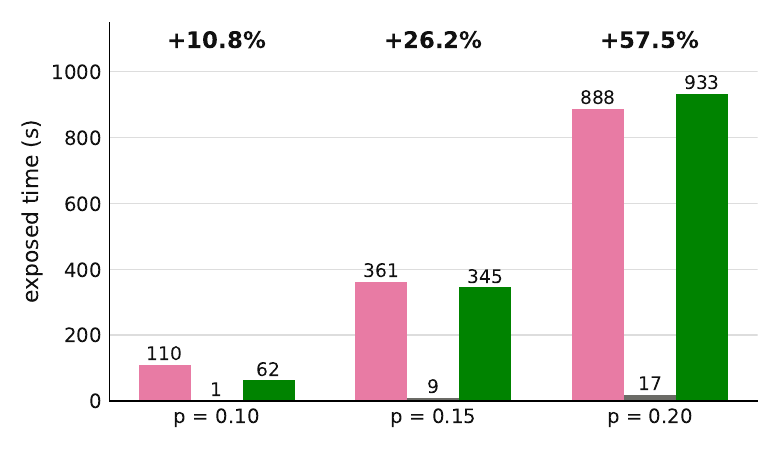}}
    \caption{QQP / RoBERTa-large}
    \label{fig:regime-large}
  \end{subfigure}
  \caption{Exposed cost components vs.\ verification rate $p$ on QQP. Bold annotations give the end-to-end overhead.}
  \label{fig:overhead-regime}
\end{figure}

\begin{figure}[t]
    \centering
    \includegraphics[width=0.8\linewidth]{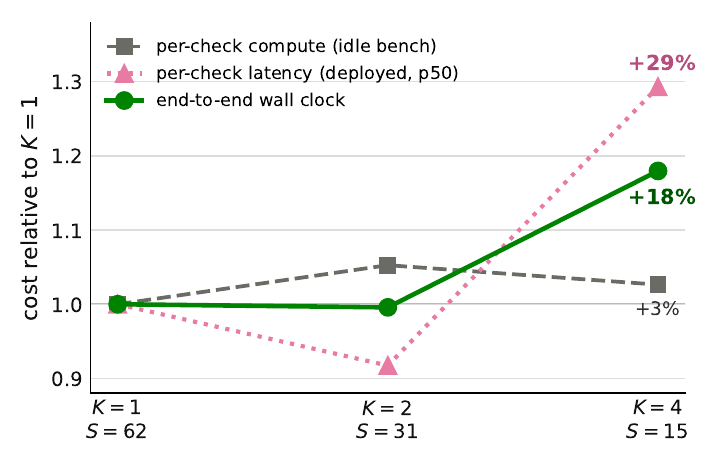}
    \caption{Verifier shape scan on the deployment pair (QQP / RoBERTa-large, $p{=}0.20$; identical audited-step set across configurations; pool size fixed at $K\!\cdot\!S\approx62$). All costs are relative to $K{=}1$.}
    \label{fig:shape-scan}
\end{figure}

We first evaluate the efficiency of our verification protocol. The main questions are: (i) how much overhead our protocol adds over ordinary single-GPU DP training, (ii) how much cost is avoided compared with the naive TEE-only design, and (iii) where the remaining cost comes from.

\smallskip \noindent \textbf{End-to-End Overhead.}
We report the results in Table~\ref{tab:overhead-breakdown}. Running the entire training loop inside the TEE is prohibitively slow, incurring $8.0\times$--$9.2\times$ overhead on the LLM tasks and $20\times$ on CIFAR-10, whereas our protocol stays below $15\%$ on every LLM task at the protocol operating point $p{=}0.10$, and below $31\%$ even at the elevated rate $p{=}0.15$. At $p{=}0.10$ the drain column is at most a few seconds on all eight tasks, indicating that almost all verification work completes by the time GPU training ends. Verification cost can nevertheless be exposed during training through synchronization and backpressure, particularly on the larger RoBERTa workloads; on the remaining tasks the residual overhead is almost entirely per-step synchronization. For example, on QQP with RoBERTa-base the end-to-end overhead is $14.5\%$, of which $47.09$\,s out of the $84.31$\,s of added wall-clock time is Sync; the RoBERTa tasks pay a visibly larger synchronization share than the GPT-2 tasks. Raising the rate to $p{=}0.15$ moves the large models toward the TEE-paced regime: on MNLI with RoBERTa-large, Sync ($477.86$\,s) and backpressure ($481.74$\,s) grow together --- the two overlap, as the training loop increasingly waits on check throughput rather than on transport --- and the overhead reaches $30.2\%$. The verification rate $p$ thus acts as a direct dial between auditing intensity and exposed cost, with the entire measured range remaining substantially cheaper than in-TEE training ($6.3\times$--$8.7\times$ across the LLM cells).

\smallskip \noindent \textbf{Overhead regimes.}
Figure~\ref{fig:overhead-regime} dissects the exposed cost as the verification rate grows, for the same task at two model sizes. Two regimes are visible. For RoBERTa-base (Fig.~\ref{fig:regime-base}), per-step synchronization dominates at every rate: checks are cheap enough ($166$ core-seconds each) that the verifier pool keeps pace with training, backpressure stays below Sync even at $p{=}0.20$, and the overhead is essentially the price of the per-step seed round-trip. For RoBERTa-large (Fig.~\ref{fig:regime-large}), whose checks cost $3.4\times$ more, the system crosses into a throughput-bound regime: backpressure grows from about half of Sync at $p{=}0.10$ ($62$\,s vs.\ $110$\,s) to overtaking it at $p{=}0.20$ ($933$\,s vs.\ $888$\,s), and much of the Sync growth in this regime is itself backpressure propagating to the GPU-side seed wait --- the training loop increasingly waits on check throughput rather than on transport. Drain stays within a few seconds in every cell: backpressure throttles training in place of letting unfinished checks spill past the epoch boundary, so verification adds at most a small tail after training completes. Larger models therefore reach the throughput-bound regime at lower verification rates, which is precisely the regime where enlarging the verifier's worker pool (rather than reducing $p$) recovers the overhead.

\smallskip \noindent \textbf{Choosing the verifier shape.}
The verifier uses $K$ concurrent checks, each split into $S$ shards, with $K\!\cdot\!S$ fixed by the available vCPUs. Figure~\ref{fig:shape-scan} shows that increasing $K$ provides little throughput benefit but increases per-check latency: at $K{=}4$, each check receives fewer workers, increasing its p50 latency by $29\%$ and the end-to-end runtime by $18\%$. A larger $K$ also keeps more verification states in flight, increasing memory pressure and queueing overhead. We therefore use small $K$ ($K\le2$) and choose $S$ per task. This result also shows that verifier shape should be selected from end-to-end measurements.

\smallskip \noindent \textbf{Additional experiments on CIFAR-10.} 
We further examine how many malicious steps are required for an effective attack on CIFAR-10. From Table \ref{tab:acc_cifar10_30_60_120}, we can see even $120$ malicious steps offer little gain at $\varepsilon=2$ or $4$, and become effective only at $\varepsilon=8$. In contrast, RoBERTa on MNLI already shows small but consistent gains with only $40$--$60$ malicious steps at $\varepsilon=2$. Hence, effective attacks on CNNs require a substantially larger $M$, which directly benefits verification efficiency.

\vspace{-0.5cm}

\section{Related Work}

\noindent \textbf{Verifiable differential privacy.}
Verifiable differential privacy certifies the correct execution of claimed
differentially private computations. Narayan et al.~\cite{narayan2015verifiable}
first introduced this concept and showed how verifiable computation can
certify differentially private data analysis. Biswas et
al.~\cite{biswas2022verifiable} applied zero-knowledge proofs to DP counting
queries, and Noisette~\cite{pang2026noisette} certifies DP noise sampling for
discrete and continuous mechanisms. More recently,
Confidential-DPProof~\cite{shahin2024confidential} and
VeriDP~\cite{veridp} use customized zero-knowledge proofs to verify
differentially private model training. These approaches provide strong
cryptographic assurance without trusted hardware, but incur substantial
proof-generation overhead for realistic training workloads. In contrast, we
explore a TEE-based design that trades complete verification for efficient
probabilistic auditing.

\smallskip \noindent \textbf{TEE-assisted verifiable machine learning.}
TEEs provide hardware-backed isolation and remote attestation, making them
useful for protecting or verifying ML computation on untrusted accelerators.
Slalom~\cite{slalom} combines a CPU-side TEE with an untrusted GPU to verify
outsourced computation. Later systems use partitioning or obfuscation to
protect sensitive model components while outsourcing most computation to the
GPU~\cite{mo2020darknetz,hou2021model,teeslice,shadownet,
zhang2024groupcover,wang2025game}. For training integrity,
TrustFL~\cite{trustfl} and GINN~\cite{ginn} use TEE-based sampled verification
of untrusted GPU computation, with GINN further combining gradient clipping
and asynchronous verification. TrustFL+~\cite{trustflplus} addresses
GPU--TEE floating-point non-determinism, while AFTUNE~\cite{aftune} uses
TEE-based spot checking for outsourced fine-tuning. In contrast, we focus on
DP training under incomplete TEE verification and explicitly quantify the
residual adversarial freedom introduced by unchecked steps and numerical
tolerance.

\smallskip \noindent \textbf{GPU TEEs and confidential GPU computing.}
Another line of work explores ML integrity through GPU TEEs~\cite{volos2018graviton,hua2020guardnn,wang2023building,shwetashinde2023acai}. However, GPU-TEE support remains limited in deployed infrastructures~\cite{wang2025game}, and many training clusters still rely on legacy GPUs. Our work is complementary: instead of requiring trusted GPUs, we target existing CPU-TEE and legacy-GPU environments.
\vspace{-0.3cm}

\section{Conclusion}
In this paper, we presented a CPU-TEE-based framework for verifiable DP training on untrusted GPUs. The protocol provides analytic accounting for arbitrary submitted aggregate gradients through probabilistic detection of full deviations and a high-probability numerical-tolerance budget. Separately, for the utility-oriented forged-gradient attacks evaluated in our experiments, sparse attack budgets provide limited utility improvement and show no measurable additional membership leakage; these empirical findings motivate the operating point used by our system rather than constituting a worst-case guarantee over all possible sparse attacks. Our evaluation shows that the design adds modest overhead over standard GPU-based DP training while avoiding the substantially higher cost of running the full training procedure inside a CPU-side TEE.

\bibliographystyle{plainurl}
\bibliography{bibfile}

\appendix

\appendix

\section{Optimizer-Aware Forged-Gradient Attack}
\label{appendix:attack-adamw}

\paragraph{Setup and threat model.} Let $\boldsymbol{\theta}_t\in\mathbb{R}^d$ denote the trainable parameters at step $t$. The optimizer is AdamW with first- and second-moment states $(\boldsymbol{m}_{t-1},\boldsymbol{v}_{t-1})$, hyperparameters $\beta_1,\beta_2\in[0,1)$, learning rate $\eta_t$, numerical constant $\epsilon$, and decoupled weight decay $\lambda$ (applied per parameter group as configured by the trainer; in the attack experiments the standard HF grouping is used, i.e.\ $\lambda=0$ for biases and LayerNorm weights). The adversary may forge the aggregated gradient $\boldsymbol{g}$ consumed by the optimizer, subject to the clipping constraint
\begin{align*}
\boldsymbol{g}\in\mathcal{B}_C
:=
\{\boldsymbol{x}:\Vert\boldsymbol{x}\Vert_2\le C\}.
\end{align*}
The realized DP noise is not observed when $\boldsymbol{g}$ is selected.

\paragraph{One-step AdamW map.} Given a candidate forged gradient $\boldsymbol{g}$, define the noise-free one-step AdamW map $\Phi_t:\mathbb{R}^d\rightarrow\mathbb{R}^d$ by
\begin{align*}
\boldsymbol{m}(\boldsymbol{g})
&=
\beta_1\boldsymbol{m}_{t-1}
+
(1-\beta_1)\boldsymbol{g},
\\
\boldsymbol{v}(\boldsymbol{g})
&=
\beta_2\boldsymbol{v}_{t-1}
+
(1-\beta_2)\boldsymbol{g}\odot\boldsymbol{g},
\\
\widehat{\boldsymbol{m}}(\boldsymbol{g})
&=
\frac{\boldsymbol{m}(\boldsymbol{g})}{1-\beta_1^t},
\qquad
\widehat{\boldsymbol{v}}(\boldsymbol{g})
=
\frac{\boldsymbol{v}(\boldsymbol{g})}{1-\beta_2^t},
\\
\Phi_t(\boldsymbol{g})
&=
\boldsymbol{\theta}_t
-
\eta_t
\left(
\frac{\widehat{\boldsymbol{m}}(\boldsymbol{g})}
{\sqrt{\widehat{\boldsymbol{v}}(\boldsymbol{g})}+\epsilon}
+
\lambda\boldsymbol{\theta}_t
\right).
\end{align*}
The current model parameters and optimizer states are fixed throughout the inner optimization.

\paragraph{Noise-free surrogate objective.} The attacker minimizes the one-step surrogate
\begin{align*}
f_t(\boldsymbol{g})
=
D\!\left(\Phi_t(\boldsymbol{g})\right),
\end{align*}
subject to $\boldsymbol{g}\in\mathcal{B}_C$.

Since our experiments fine-tune LoRA parameters and, for classification tasks, an additional classification head, we define the steering distance as
\begin{align*}
D(\boldsymbol{\theta})
=
\sum_{\text{LoRA pairs }j}
\left\|
s_j\boldsymbol{B}_j\boldsymbol{A}_j
-
s_j\boldsymbol{B}_j^\star\boldsymbol{A}_j^\star
\right\|_F^2
+
\sum_{\text{unpaired }i}
\left\|
\theta_i-\theta_i^\star
\right\|_2^2,
\end{align*}
where $s_j=\alpha_j/r_j$ is the LoRA scaling.

For LoRA parameters, the distance is measured on the merged update $\Delta W_j=s_jB_jA_j$ rather than directly on the factors $(A_j,B_j)$. This removes the ambiguity caused by the non-unique LoRA factorization: for any invertible $G\in\mathbb{R}^{r_j\times r_j}$, the reparameterization $(\boldsymbol{A}_j^\star,\boldsymbol{B}_j^\star)\mapsto(G\boldsymbol{A}_j^\star,\,\boldsymbol{B}_j^\star G^{-1})$ leaves the merged update unchanged,
\begin{align*}
(\boldsymbol{B}_j^\star G^{-1})(G\boldsymbol{A}_j^\star)
=
\boldsymbol{B}_j^\star\boldsymbol{A}_j^\star ,
\end{align*}
so a distance taken on the factors would charge for the choice of $G$ while a distance taken on the product does not. Consequently, the attack objective depends only on the target merged update rather than on a particular choice of target factors.

\subsection{Solving the Inner Problem}

\paragraph{Linearized initialization.} We initialize the attack using the constrained minimizer of the first-order expansion of $f_t$ around $\boldsymbol{g}=\boldsymbol{0}$. Let
\begin{align}
\boldsymbol{w}_t
:=
\nabla_{\boldsymbol{g}}f_t(\boldsymbol{g})
\big|_{\boldsymbol{g}=\boldsymbol{0}}.
\label{eq:attack-gradient}
\end{align}
Then the minimizer of the linearized objective over $\mathcal{B}_C$ is
\begin{align*}
\boldsymbol{g}^{(0)}
=
-C\frac{\boldsymbol{w}_t}{\Vert\boldsymbol{w}_t\Vert_2}.
\end{align*}

The gradient in \eqref{eq:attack-gradient} is obtained by automatic differentiation through the complete surrogate $D(\Phi_t(\boldsymbol{g}))$. Equivalently,
\begin{align*}
\boldsymbol{w}_t
=
\boldsymbol{J}_0^\top
\nabla_{\boldsymbol{\theta}}
D\!\left(\Phi_t(\boldsymbol{0})\right),
\qquad
\boldsymbol{J}_0
=
\frac{\partial\Phi_t}{\partial\boldsymbol{g}}
\Big|_{\boldsymbol{g}=\boldsymbol{0}},
\end{align*}
where $\boldsymbol{J}_0$ is diagonal, so the product above is elementwise.
In particular, the steering gradient is evaluated at $\Phi_t(\boldsymbol{0})$, rather than at $\boldsymbol{\theta}_t$, since the existing AdamW first moment and weight decay can move the parameters even when the newly submitted gradient is zero.

Because \eqref{eq:attack-gradient} differentiates the same merged-space objective used by the attack, the initialization is independent of the particular factorization used to represent the target LoRA adapters.

\paragraph{Projected gradient refinement.} Starting from $\boldsymbol{g}^{(0)}$, we refine the forged gradient using projected gradient descent on the exact surrogate:
\begin{align*}
\boldsymbol{g}^{(k+1)}
=
\Pi_{\mathcal{B}_C}
\left(
\boldsymbol{g}^{(k)}
-
\gamma_k C
\frac{
\nabla_{\boldsymbol{g}}f_t(\boldsymbol{g}^{(k)})
}{
\Vert\nabla_{\boldsymbol{g}}f_t(\boldsymbol{g}^{(k)})\Vert_2
}
\right),
\end{align*}
where
\begin{align*}
\Pi_{\mathcal{B}_C}(\boldsymbol{x})
=
\boldsymbol{x}
\min\left\{
1,
\frac{C}{\Vert\boldsymbol{x}\Vert_2}
\right\}.
\end{align*}
We use backtracking line search: a candidate is accepted only if it is finite and decreases $f_t$; otherwise the step size is halved and retried. On acceptance the step size is doubled for the next iteration, capped at $\gamma$, so the search can recover a large step after a locally difficult region. The optimization terminates after at most $K$ iterations, when no backtracking step decreases the objective, or when the change in $\boldsymbol{g}$ falls below tolerance $\tau$.

\paragraph{Noisy execution.} Let $\boldsymbol{g}^\star$ denote the forged gradient returned by the inner optimization. The actual optimizer consumes
\begin{align*}
\widetilde{\boldsymbol{g}}
=
\boldsymbol{g}^\star
\min\left\{
1,
\frac{C}{\Vert\boldsymbol{g}^\star\Vert_2}
\right\}
+
\frac{1}{B}\boldsymbol{\xi},
\qquad
\boldsymbol{\xi}
\sim
\mathcal{N}
\left(
\boldsymbol{0},
\sigma^2C^2\boldsymbol{I}
\right).
\end{align*}
Thus the attack is optimized using the noise-free surrogate, while the realized AdamW update and optimizer states are formed from the noisy gradient $\widetilde{\boldsymbol{g}}$.

$\eta_t$ is taken from the live
parameter group before the scheduler has written the current step's value, so
the surrogate sees $\eta_{t-1}$. It cancels from the solution almost entirely:
$\eta_t$ multiplies the whole Jacobian and so leaves any
normalised direction unchanged, surviving only through the point $\Phi_t(\boldsymbol{0})$
at which $\nabla D$ is read. Nor does that residue grow late in training --- under
a linear decay $\eta_t-\eta_{t-1}=\eta_0/T$ is constant while
$\Phi_t(\boldsymbol{0})\to\boldsymbol{\theta}_t$. Rebuilding the seed at both rates, each with
$\nabla D$ re-evaluated at its own $\Phi_t(\boldsymbol{0})$, turns its direction by
$1-\cos=1.2\times10^{-14}$ (QQP/roberta-base) and $5.7\times10^{-14}$
(E2E/gpt-2), both at $t=0.6\,T$ and for the seed alone --- seven orders below the
fp32 precision the deployed solver runs at. It is recorded because ``read from
the live optimizer state'' would otherwise be read as exact.

\section{Description of Used Datasets}
\label{appendix:data_description}

We evaluate our method on datasets covering tabular classification, image classification, natural language understanding (NLU), and natural language generation (NLG) tasks.

\noindent\textbf{Purchase~\cite{purchase}.} Purchase is a tabular classification dataset widely used in privacy and membership-inference studies. Each record represents a user's purchase behavior, and the goal is to predict the corresponding purchase category. We use this dataset to evaluate DP training on multilayer perceptrons (MLPs), which serve as representative models for conventional tabular learning tasks.

\noindent\textbf{CIFAR-10~\cite{krizhevsky2009learning}.} CIFAR-10 is a standard image classification benchmark consisting of natural images from ten object categories. We use CIFAR-10 to evaluate DP training on convolutional neural networks (CNNs), where the model is trained to classify each image into one of the ten classes. The model is a WideResNet-16-4 trained on $40{,}000$ target training images, with $15{,}000$ images reserved as shadow data for the membership-inference experiments and $5{,}000$ as the test set; the honest DP-SGD runs and the attacked runs share the same recipe: $\varepsilon=2$ (Table~\ref{tab:acc_cifar10_30_60_120} additionally reports $\varepsilon=4$ and $8$), $\delta=10^{-5}$, $5{,}000$ training steps, batch size $2{,}048$ with microbatches of $512$, learning rate $2.0$, and clipping norm $C=1$.

\noindent\textbf{MNLI~\cite{williams2018broad}.} The Multi-Genre Natural Language Inference (MNLI) dataset is a natural language understanding benchmark. Given a premise and a hypothesis, the task is to predict whether the hypothesis is entailed by, contradicts, or is neutral with respect to the premise. We use MNLI to evaluate DP fine-tuning for sentence-pair understanding tasks.

\noindent\textbf{QQP~\cite{wang2018glue}.} The Quora Question Pairs (QQP) dataset is another natural language understanding benchmark. Given a pair of questions, the task is to determine whether the two questions are semantically equivalent. We use QQP to evaluate DP fine-tuning on semantic matching and paraphrase detection tasks.

\noindent\textbf{E2E~\cite{novikova2017e2e}.} The E2E dataset is a natural language generation benchmark for data-to-text generation. Given a structured meaning representation, the task is to generate a fluent natural-language utterance that accurately describes the input attributes. We use E2E to evaluate DP fine-tuning for controlled text generation.

\noindent\textbf{WebNLG~\cite{gardent2017webnlg}.} WebNLG is a data-to-text natural language generation dataset built from structured RDF triples. Given a set of subject--predicate--object triples describing entities and relations, the model must generate a coherent natural-language description that preserves the input facts. Compared with slot-based datasets such as E2E, WebNLG covers more diverse domains and relational structures, making it useful for evaluating DP fine-tuning on structured generation tasks.

\subsection{Attack Configuration}
\label{appendix:attack-config}

Table~\ref{tab:attack-params} lists the per-setting constants.\footnote{We use the
\texttt{dp-transformers} trainer~\cite{dp-transformers}: fixed-size batches drawn
from a per-epoch reshuffle with the last partial batch dropped, and RDP/PRV
accounting at sampling rate $B/N$. Such means has been acknowledged as a common
practice~\cite{chua2024shuffling}.} The
remaining choices are shared across all eight settings and are stated here.

\noindent\textbf{Budgets and strategies.} The deviation budget is
$M\in\{40,50,60\}$ and each budget is run under all three target strategies
(final, nearest, current), giving $3\times3\times8=72$ attacked runs alongside
the eight honest DP baselines and the eight non-private references.

\noindent\textbf{Which steps are deviated on.} The adversary chooses the step set
rather than drawing it at random, so the reported numbers are its best placement.
Both schedules concentrate on $u^\star$, the fraction of training at which one
unit of deviation buys the most displacement at the end. The final-checkpoint
attack, whose target is fixed, injects on the $M$ \emph{consecutive} steps centred
on $u^\star T$. The nearest- and current-checkpoint attacks instead take
one step from each window of width $T/M$, at the point of that window closest to
$u^\star T$. We did not run a uniform-random control.

\noindent\textbf{Inner loop.} The forged gradient is refined by projected
gradient descent from the linearised seed: step size $\gamma=1$ as a fraction of
the clipping norm $C$, at most $K=100$ iterations, up to ten halvings of the trial
step per iteration, and a stopping tolerance of $10^{-6}$ on
$\lVert\boldsymbol{g}^{(k+1)}-\boldsymbol{g}^{(k)}\rVert_2$. A candidate is accepted only if it is
finite and decreases the surrogate.

\noindent\textbf{Non-private reference.} For the classification tasks the target
is the final non-private checkpoint. For E2E and WebNLG\footnote{On the
generation tasks the per-example loss is the token mean over that example's own
unmasked positions and the backward scalar is the mean over examples, so each
per-example gradient is the gradient of one record's loss rather than of a
batch-wide token mean.} it is
the non-private checkpoint with the best validation score, which is not the last one in three of
the four settings; the column in Table~\ref{tab:attack-params} names the
checkpoint used in each case.

\begin{table*}[t]
\centering
\small
\setlength{\tabcolsep}{4pt}
\caption{Parameters of the forged-gradient experiments. All eight settings share $\varepsilon=2$, clipping norm $C=1$, learning rate $4e-4$ with a linear schedule and no warm-up, weight decay $10^{-2}$, AdamW $(\beta_1,\beta_2)=(0.9,0.999)$, $\epsilon_{\mathrm{Adam}}=10^{-8}$, 10 epochs, seed 1, and LoRA dropout $0$ with every other dropout module zeroed. $T$ is the number of optimizer steps (the sampler drops the last partial batch, so $T=\lfloor N/B\rfloor\times$epochs). $\sigma$ is the noise multiplier the accountant returns at that $(\varepsilon,\delta,B/N,T)$; the per-coordinate noise actually added is $\sigma C/B$. $u^\star$ is the centre of the injection window as a fraction of training. The target is the non-private checkpoint the adversary steers towards under the \emph{final} strategy.}
\label{tab:attack-params}
\begin{tabular}{llrrrrrrrrrl}
\toprule
task & model & $N$ & $B$ & $T$ & seq & $r$ & $\alpha$ & $\delta$ & $\sigma$ & $\sigma C/B$ & $u^\star$ / target \\
\midrule
QQP & roberta-base & 363,846 & 256 & 14,210 & 128 & 8 & 8 & $10^{-6}$ & 0.6462 & 0.00252 & 0.20 / \texttt{checkpoint-14220} \\
QQP & roberta-large & 363,846 & 256 & 14,210 & 128 & 8 & 8 & $10^{-6}$ & 0.6462 & 0.00252 & 0.20 / \texttt{checkpoint-14220} \\
MNLI & roberta-base & 392,702 & 256 & 15,330 & 128 & 8 & 8 & $10^{-6}$ & 0.6407 & 0.00250 & 0.20 / \texttt{checkpoint-15340} \\
MNLI & roberta-large & 392,702 & 256 & 15,330 & 128 & 8 & 8 & $10^{-6}$ & 0.6407 & 0.00250 & 0.20 / \texttt{checkpoint-15340} \\
E2E & gpt2 & 42,061 & 64 & 6,570 & 128 & 4 & 8 & $10^{-5}$ & 0.6679 & 0.01044 & 0.10 / \texttt{checkpoint-3290} \\
E2E & gpt2-medium & 42,061 & 64 & 6,570 & 128 & 4 & 8 & $10^{-5}$ & 0.6679 & 0.01044 & 0.10 / \texttt{checkpoint-5922} \\
WebNLG & gpt2 & 18,025 & 32 & 5,630 & 256 & 4 & 8 & $10^{-5}$ & 0.6827 & 0.02133 & 0.10 / \texttt{checkpoint-5076} \\
WebNLG & gpt2-medium & 18,025 & 32 & 5,630 & 256 & 4 & 8 & $10^{-5}$ & 0.6827 & 0.02133 & 0.10 / \texttt{checkpoint-5640} \\
\bottomrule
\end{tabular}

\end{table*}

\section{Generation Metrics Beyond BLEU}
\label{appendix:aux-metrics}

Figure~\ref{fig:grouped_bar} reports BLEU for E2E and WebNLG. Both scorers emit
further metrics, and since a single metric can move for reasons unrelated to
output quality we report all of them here. Table~\ref{tab:aux-metrics} gives, for
each setting and each metric, the DP baseline, the most favourable of the nine
attacked runs \emph{for that metric}, and the non-private reference. Taking the
best cell per metric rather than per run is deliberate: it asks whether any
attacked run moves the metric at all, which is the more demanding question.

No metric shows a substantial absolute gain. On E2E/gpt2-medium, where the
attacked runs do move every metric in the same direction as BLEU, the largest
changes are $+0.014$ METEOR, $+0.113$ NIST, $+0.081$ CIDEr and $+0.005$
ROUGE-L --- $0.7$ to $4.1$ percent of the metric's own value, against
non-private references that remain $0.04$ to $0.33$ away. On E2E/gpt2 the metrics
do not even agree in sign: NIST and METEOR rise while BLEU, ROUGE-L and CIDEr
fall, which is what one expects when the effect is smaller than the noise of
selecting a decoding configuration on a $547$-example validation split, and we
therefore draw no conclusion from that setting. On both WebNLG settings every
non-BLEU change other than TER is zero or a single unit in the last printed
digit, and TER improves by $0.02$ and $0.04$; the evaluator prints those
metrics to two decimals, so they cannot resolve an effect of this size in
either direction.

The reading consistent with all four settings is that sparse deviations do not
produce a clear gain on any generation metric. Where the metrics are precise
enough to resolve the effect at all, the absolute movement is small; where they
are not, they are silent rather than supportive.

\begin{table*}[t]
\centering
\small
\setlength{\tabcolsep}{5pt}
\caption{Every metric the official scorers emit, on the test split at $\varepsilon=2$. ``best attack'' is the most favourable of the nine attacked runs \emph{for that metric}, so different rows may be won by different $(M,\text{strategy})$ cells; $\Delta$ is its absolute change from the DP baseline, in the metric's own units. TER is lower-is-better, so a negative $\Delta$ is an improvement there. The WebNLG evaluator prints every metric but BLEU to two decimals, which is the granularity of the changes observed on that task. At that granularity the best value is usually reached by several cells at once, and the last column then reports how many rather than naming one.}
\label{tab:aux-metrics}
\begin{tabular}{llrrrrl}
\toprule
setting & metric & DP & best attack & $\Delta$ & non-private & best cell \\
\midrule
E2E / gpt2 & BLEU & 0.6044 & 0.5999 & -0.0045 & 0.6495 & $M{=}$40, current \\
 & NIST & 7.8193 & 7.9538 & +0.1345 & 8.2836 & $M{=}$40, current \\
 & METEOR & 0.3833 & 0.3906 & +0.0073 & 0.4219 & $M{=}$60, final \\
 & ROUGE\_L & 0.6211 & 0.6156 & -0.0055 & 0.6726 & $M{=}$60, nearest \\
 & CIDEr & 1.8375 & 1.8128 & -0.0247 & 2.1242 & $M{=}$60, current \\
\midrule
E2E / gpt2-medium & BLEU & 0.6144 & 0.6230 & +0.0086 & 0.6677 & $M{=}$40, nearest \\
 & NIST & 8.0637 & 8.1765 & +0.1128 & 8.5079 & $M{=}$40, final \\
 & METEOR & 0.4087 & 0.4224 & +0.0137 & 0.4608 & $M{=}$50, final \\
 & ROUGE\_L & 0.6563 & 0.6609 & +0.0046 & 0.7015 & $M{=}$50, nearest \\
 & CIDEr & 1.9659 & 2.0464 & +0.0805 & 2.3245 & $M{=}$50, final \\
\midrule
WebNLG / gpt2 & BLEU & 34.55 & 35.26 & +0.71 & 46.82 & $M{=}$50, final \\
 & BLEU\_NLTK & 0.3300 & 0.3400 & +0.0100 & 0.4600 & 9 of 9 cells tied \\
 & METEOR & 0.3100 & 0.3200 & +0.0100 & 0.3800 & 9 of 9 cells tied \\
 & chrF++ & 0.5300 & 0.5400 & +0.0100 & 0.6400 & 4 of 9 cells tied \\
 & TER & 0.6100 & 0.5900 & -0.0200 & 0.5000 & 4 of 9 cells tied \\
\midrule
WebNLG / gpt2-medium & BLEU & 38.53 & 39.51 & +0.98 & 49.39 & $M{=}$60, final \\
 & BLEU\_NLTK & 0.3800 & 0.3800 & +0.0000 & 0.5000 & 9 of 9 cells tied \\
 & METEOR & 0.3500 & 0.3500 & +0.0000 & 0.4000 & 3 of 9 cells tied \\
 & chrF++ & 0.5900 & 0.5900 & +0.0000 & 0.6700 & 2 of 9 cells tied \\
 & TER & 0.5900 & 0.5500 & -0.0400 & 0.4300 & 5 of 9 cells tied \\
\bottomrule
\end{tabular}
\end{table*}

\section{Analogy to Learning with Bounded Adversarial Gradient Perturbations}
\label{appendix:bounded_gradient_perturbation}

We further discuss a related theoretical perspective from distributed learning with bounded adversarial gradient perturbations. This line of work considers a setting where a learner queries gradients from clients, but each returned gradient may be adversarially perturbed subject to a bounded-distance constraint. More concretely, for a client loss function $\ell_i$, the returned vector $u$ need not equal the true gradient $\nabla \ell_i(w)$; it only needs to satisfy
\begin{align*}
\Vert
u-\nabla \ell_i(w)
\Vert_2
\le
\epsilon.
\end{align*}
Under convex and $L$-smooth objectives, the study shows that such adversarial perturbations do not make learning arbitrary. Instead, when an upper bound $\Vert w^\star\Vert_2\le R$ on the optimal solution is known, the achievable optimization error is controlled by the perturbation radius and problem parameters: the minimum unavoidable sub-optimality is on the order of $\epsilon R$, and an algorithm can guarantee a sub-optimality gap of order $\epsilon R$ with finite query complexity.

Although our setting is different, this result provides a useful analogy for understanding why sparse malicious deviations in our protocol provide limited utility benefit. In the deployed protocol, per-example clipping and aggregation are executed on the untrusted GPU, but every submitted aggregate is subject to the always-on norm invariant $\Vert\bar g_t^{\mathrm{GPU}}\Vert_2\le C(1+10^{-4})$, while the prescribed honest clipped aggregate $\bar g_t^{\mathrm{ref}}$ has norm at most $C$. By the triangle inequality,
\begin{align*}
\Vert\bar g_t^{\mathrm{GPU}}-\bar g_t^{\mathrm{ref}}\Vert_2
\le
\Vert\bar g_t^{\mathrm{GPU}}\Vert_2
+
\Vert\bar g_t^{\mathrm{ref}}\Vert_2
\le
(2+10^{-4})C.
\end{align*}
Hence every accepted submission lies within $(2+10^{-4})C$ of the trusted aggregate. This provides a bounded-perturbation analogy, analogous to the bounded adversarial-gradient model above, where each adversarial reply may deviate from the true gradient only within a finite radius; it should not be interpreted as the formal security theorem of our non-convex stochastic training setting.

This analogy supports the intuition that the norm invariant limits the adversary's per-step influence. The adversary may still bias the optimization trajectory, but it cannot inject an unbounded update through a single gradient reply.

\noindent\textbf{Why the related result is more general in adversarial strength?} The bounded-perturbation setting considered in the prior theoretical study is more general than our tolerated sparse-deviation regime in several important ways. First, it allows every gradient query to be adversarially perturbed, as long as the returned vector remains within the prescribed perturbation radius. In contrast, our relaxed security goal only tolerates sparse deviations: the adversary may deviate on a small fraction of iterations, while repeated full deviations are detected with high probability by our probabilistic checking mechanism. Second, the adversarial perturbation in that model is worst-case and can be chosen adaptively at each query, whereas in our protocol the submitted aggregate is subject to the always-on aggregate-norm invariant and to the trusted DP-SGD update rule, while GPU-side clipping and aggregation are probabilistically checked by trusted recomputation. Third, our training process additionally includes DP noise injected inside the TEE, which is not controlled by the trainer and further limits the trainer's ability to precisely steer the final model.

Therefore, the prior result should not be interpreted as a direct proof for our non-convex, stochastic DP training setting. Nevertheless, it gives a useful conceptual justification: even in a stronger setting where the adversary can perturb every gradient within a bounded radius, the resulting optimization error remains controlled rather than arbitrary. Since our tolerated adversary is further restricted to sparse deviations, the cumulative benefit in our setting is intuitively even more limited. The related result therefore provides a useful bounded-perturbation analogy rather than a formal guarantee for our non-convex stochastic setting. In our protocol, clipping and the aggregate-norm invariant constrain the magnitude of each submitted update, while the limited utility impact of sparse deviations is established empirically for the attack family evaluated in Section~\ref{section:analysis_on_malicious_behavior}.

\section{Calibration and Security Cost of Numerical Tolerance}
\label{appendix:numerical-calibration}

This appendix complements Section~\ref{section:gpu-tee-discrepancy} with the precise definitions, the calibration rules, and the formal accounting behind the numerical-tolerance budget.

For a checked step, define
\begin{align*}
z_t^{32}
&=
\frac{
\Vert\bar g_t^{\mathrm{GPU}}-\bar g_t^{32}\Vert_2
}{C},
&
z_t^{64}
&=
\frac{
\Vert\bar g_t^{\mathrm{GPU}}-\bar g_t^{64}\Vert_2
}{C}.
\end{align*}
The threshold $\tau_{\mathrm{abs}}$ routes a step to FP64 adjudication. An escalated step is accepted only if
\begin{align*}
z_t^{64}
\le
\rho_{\mathrm{amb}},
\qquad
\rho_{\mathrm{amb}}
=
\frac{2\tau_{\mathrm{num}}}{B}.
\end{align*}
Thus, $\rho_{\mathrm{amb}}$ is the security-relevant aggregate acceptance radius; $\tau_{\mathrm{num}}$ is only its implementation-specific normalization.

\subsection{Honest Calibration and False Abort}

For a fixed honest trajectory $r$, partition the steps into
\begin{align*}
\mathcal U_r
&=
\{t:z_{t,r}^{32}\le\tau_{\mathrm{abs}}\},
\\
\mathcal A_r
&=
\{t:z_{t,r}^{32}>\tau_{\mathrm{abs}},
      z_{t,r}^{64}\le\rho_{\mathrm{amb}}\},
\\
\mathcal H_r
&=
\{t:z_{t,r}^{32}>\tau_{\mathrm{abs}},
      z_{t,r}^{64}>\rho_{\mathrm{amb}}\}.
\end{align*}
Let
\begin{align*}
M_r=|\mathcal A_r|,
\qquad
H_r=|\mathcal H_r|,
\end{align*}
and define the sampled body charge
\begin{align*}
Z_r
=
\sum_{t\in\mathcal U_r}
V_tz_{t,r}^{32},
\qquad
V_t\sim\operatorname{Bernoulli}(p).
\end{align*}
Writing
\begin{align*}
F_{\mathrm{sub}}^{(r)}(x)
=
\Pr[Z_r\le x],
\end{align*}
and conditioned on the honest trajectory satisfying the always-on aggregate-norm invariant (observed on all calibration trajectories and the held-out replay), the exact trajectory-conditional false-abort probability under independent Bernoulli checking is
\begin{align}
q_r^{\mathrm{FA}}
=
1
-
F_{\mathrm{sub}}^{(r)}(K_{\mathrm{sub}})
F_{\mathrm{Bin}}(K_{\mathrm{amb}};M_r,p)
(1-p)^{H_r}.
\label{eq:false-abort-numerical}
\end{align}

Equation~\eqref{eq:false-abort-numerical} is conditional on the realized honest trajectory and makes no stationarity assumption on numerical errors.

\paragraph{Sampling laws in calibration and deployment.}
The deployed protocol uses independent hidden verification coins
$V_t\sim\operatorname{Bernoulli}(p)$, and all security and false-abort
probabilities reported in this paper are evaluated under this law. The
numerical data used to select and validate the verifier parameters were
collected through several measurement procedures: (i) the broad
calibration campaign, $32$ trajectories on a calibration GPU--TEE pair,
carries $z^{64}$ on every step and $z^{32}$ on the harness's uniformly random
audits (rate $0.1$) plus probed tail steps. (ii) The deployment-pair systematic traces, eight honest
pilot trajectories on the target pair, carry every-tenth-step $z^{32}$ and
full-step FP64 records; seven of them
(all but qqp-large) are also the trajectories on which the family-level tail
values of Table~\ref{table:verifier-params} are evaluated. (iii) The qqp-large replay, a trajectory
preregistered with a fresh seed after parameter freezing and recorded on
every step, is the only held-out trajectory and hence the only false-abort
estimate; and (iv) the overhead harness fixes a hidden uniformly
random audit subset of size $\lceil pT\rceil$ in order to stabilize the
verification workload during timing measurements. Because honest
verification is passive (the audited recomputation does not alter the
trajectory; replay fidelity was verified bitwise), the underlying trajectory
statistics $(S_r,Q_r,M_r,H_r)$ do not depend on the audit law; the
observation procedure determines only how completely those statistics are
measured. For the systematic traces of (ii) we treat the every-tenth-step
sample as representative of the trajectory when forming the trajectory-level
upper estimates $S_r^{\mathrm{up}},Q_r^{\mathrm{up}},M_r^{\mathrm{up}}$ (no
period-ten structure in the numerical discrepancy); the resulting values are
model-based estimates under this representativeness assumption rather than
distribution-free confidence bounds, and on the census trajectory the
estimates obtained from each of the ten offsets cover the exact statistics.
The hard-rejection count $H_r$ is exact on every trajectory and does not rely
on the assumption. For CIFAR-10 the statistics come from the harness's
audited steps, a uniformly random subset for which the binomial count
inversion is conservative.
Table~\ref{table:data-provenance} summarizes the data sources.

\paragraph{Staged calibration.} The verifier parameters are obtained through the calibration procedure developed from the broad numerical-discrepancy campaign and subsequently instantiated on the target GPU--TEE pair. The broad campaign characterizes the body/tail structure, fallback behavior, and cross-run variation needed to establish the calibration rules; its four runs per configuration share the data order and differ only in a shifted noise stream, so the cross-run variation it exhibits understates that of independent replicates, and no reported quantity relies on it beyond the $+20\%$ sensitivity row. Before certified deployment, we apply these rules to honest pilot trajectories on the target pair, consolidate the resulting values at the model-family level, and freeze the complete verifier configuration. A change in GPU family or in the trusted reference stack triggers recalibration; a change in the GPU-side training stack is revalidated against the calibrated discrepancy envelope and triggers recalibration if the observed honest spectrum falls outside it.

\paragraph{Environment differences between the calibration stages and the deployed pair.} Not all calibration trajectories were produced under the deployed GPU software stack. For the RoBERTa-large configurations the GPU side of both calibration stages ran under the deployed stack; for RoBERTa-base and the four GPT-2 configurations it ran under an earlier PyTorch build and a PEFT version whose LoRA initialisation differs, so those calibration trajectories belong to a different honest-trajectory family than the deployed runs. The TEE-side reference is unaffected, because every reference recomputation copies the trainable state from the GPU-side record. The fallback radius and the escalation-count inputs, which the procedure takes from the broad campaign, were checked on the deployed pair: the full-step FP64 census of the eight deployment-pair traces and of the held-out replay stays below $\rho_{\mathrm{amb}}$ (largest $7.81\times10^{-3}$ against $1.185\times10^{-2}$ for RoBERTa, $6.58\times10^{-4}$ against $4.891\times10^{-3}$ for GPT-2, so $H_r=0$ exactly), and the escalation counts observed on the deployed pair (at most two on any every-tenth-step trace, $22$ on the full-census replay) stay below the campaign-derived opportunity bounds and far below $K_{\mathrm{amb}}$. The deployment runs of Section~\ref{section:experiment} likewise use a GPU-side software stack that differs from the deployment-pair calibration trajectories, so we revalidate the frozen configuration against their honest discrepancy spectrum: over the $3770$ checks judged in those runs, the ratio of the observed body-discrepancy means to those of the calibration traces lies in $0.81$--$1.23$ for six of the eight tasks and below $0.81$ for the remaining two (the conservative direction), with a single escalation, accepted by the FP64 check ($n_t=0.91\le\tau_{\mathrm{num}}$), and no norm violation; this lies inside the margins of Table~\ref{table:verifier-params} and within the $+20\%$ sensitivity row of the evaluation paragraph, so we retain the frozen parameters for these runs. The equivalence of the TEE reference itself was tested once rather than assumed: recomputing recorded calibration steps of both families on the host reference environment and inside the SEV-SNP guest gave bitwise-identical discrepancies, and recomputing one recorded step with the verifier's shard grouping (micro-batches of $256$, $8$, and $4$ examples; $16$ and $1$ threads) reproduced the recorded value bitwise, so the host-side calibration passes are taken as the deployed TEE computation. The FP64 references of the calibration stages were computed on a GPU, whereas the deployed verifier computes them on the TEE CPU; recomputing four recorded steps on the TEE CPU (two from the campaign and two from the deployment-pair traces, including the two largest FP64 discrepancies observed, $n_t\approx1.0$) reproduced the GPU values to within $3\times10^{-9}$ relative, an absolute difference in $z^{64}$ below $10^{-12}$, ten orders of magnitude below $\rho_{\mathrm{amb}}$, which we treat as negligible. For CIFAR-10 the calibration and deployment runs share one GPU environment and one TEE environment; in a small-scale test on a single audited step, host and guest recomputations agreed bitwise, and recomputing that step with chunk sizes of $2048$, $256$, $33$ (the verifier's shard width), and $8$ examples changed $z$ by at most $10^{-7}$ relative (fp32 summation order; not bitwise), three orders of magnitude below $\tau_{\mathrm{abs}}$; on this small-scale evidence we treat the effect as negligible.

\paragraph{In-sample calibration checks and the held-out estimate.} The family-level values of Table~\ref{table:verifier-params} are evaluated on seven of the eight honest trajectories used during deployment-pair calibration (all but qqp-large; see the evaluation paragraph), the same trajectories on which the calibration procedure was instantiated. They confirm that the frozen parameters meet the design target on those trajectories, under the corresponding evidence model, and are therefore in-sample calibration checks. A false-abort estimate is available for exactly one deployment trajectory: after the parameters were frozen we preregistered a new qqp-large trajectory with a fresh seed, ran it on the target pair under the deployed software stack, and replayed it with full-step numerical recording; this held-out replay passed the verifier with a trajectory-conditional false-abort upper bound of $4.1\times10^{-100}$ and retuned no parameter.

\begin{table}[t]
\centering
\footnotesize
\caption{Provenance of the honest numerical data.}
\label{table:data-provenance}
\begin{tabular}{@{}p{0.33\columnwidth}p{0.37\columnwidth}p{0.22\columnwidth}@{}}
\toprule
data source & observation pattern & role \\
\midrule
broad calibration campaign ($32$ traj., calibration pair) & $z^{64}$ every step; $z^{32}$ on random audits $+$ probes & reference characterization \\
deployment-pair traces ($8$ traj., target pair) & every-tenth-step $z^{32}$; full-step FP64 & deployment-pair calibration; in-sample check \\
qqp-large preregistered replay (target pair) & every step & held-out evaluation \\
CIFAR-10 traces ($4$ traj., target pair) & $z$ on harness audits (random subset) & deployment-pair calibration; in-sample check \\
\bottomrule
\end{tabular}
\end{table}

\paragraph{Evaluating Equation~\eqref{eq:false-abort-numerical}.} For a trajectory with a full census, $M_r$ and $H_r$ are read directly, and for the fully observed body sequence $\{z_{t,r}^{32}\}_{t\in\mathcal U_r}$ the independence of the Bernoulli checking coins gives the Chernoff bound
\begin{align}
1-F_{\mathrm{sub}}^{(r)}(K_{\mathrm{sub}})
\le
\inf_{\lambda>0}
\exp(-\lambda K_{\mathrm{sub}})
\prod_{t\in\mathcal U_r}
\left(
1-p+p\,e^{\lambda z_{t,r}^{32}}
\right).
\label{eq:chernoff-body}
\end{align}
When the body was recorded on a systematic sample (every tenth step), the realized moments are replaced by trajectory-level upper estimates computed as nominal simultaneous $95\%$ confidence bounds for a uniform random sample of the same size; these are model-based estimates under the assumption that the systematic sample is representative of the trajectory (see the sampling-laws paragraph above), not distribution-free confidence bounds. For CIFAR-10, where discrepancies are recorded only on the harness's randomly audited steps, the honest escalation rate is instead bounded by the rule of three on the calibration audits, $F_{\mathrm{Bin}}(K_{\mathrm{amb}};M_r,p)$ is replaced by the corresponding binomial bound over the ${\approx}500$ checked steps of the $5{,}000$-step run at $p{=}0.1$, and the hard-cap term $(1-p)^{H_r}$ is supported empirically only (no honest audited step exceeded $\rho_{\mathrm{ill}}$ during calibration). Within the corresponding evidence model, each family-level value in Table~\ref{table:verifier-params} is at least the largest evaluated per-trajectory upper value within that family. The qqp-large calibration trace is excluded from the RoBERTa entry: under the systematic-sample evidence model the range term of its body estimate leaves the bound at about $0.2$, far above the target, which is why that configuration was re-examined by the full-census replay. As a sensitivity check rather than a bound, inflating every calibrated trajectory statistic ($S_r^{\mathrm{up}}$, $Q_r^{\mathrm{up}}$, $M_r^{\mathrm{up}}$, and the CIFAR-10 escalation rate) by $20\%$ gives worst-case conditional values of $1.3\times10^{-6}$ (RoBERTa), $4\times10^{-4}$ to $1\times10^{-3}$ (GPT-2, depending on whether the second-moment estimate is inflated linearly or quadratically), and $2.9\times10^{-4}$ (CIFAR-10); a $30\%$ inflation exceeds the $10^{-3}$ target for GPT-2. The thinnest margin is the GPT-2 body charge, $pS_r^{\mathrm{up}}/K_{\mathrm{sub}}=0.73$ on the worst trajectory (E2E, GPT-2-medium), against at most $0.43$ for RoBERTa.

\subsection{Security Accounting}
\label{appendix:security_accounting}
\paragraph{Exhaustive classification of GPU submissions.}
For every committed GPU submission that satisfies the always-on structural checks, the trusted recomputation defines a counterfactual verification outcome independently of whether that step is actually sampled. The submission therefore belongs to exactly one of three disjoint classes: the FP32 body $\mathcal B$, the accepted FP64 ambiguity region $\mathcal A$, or the hard-rejection region $\mathcal H$. The first two classes constitute the numerical-tolerance channels and are accounted for by $(G_{\mathrm{extra}},\beta_{\mathrm{extra}})$ below. Every step in $\mathcal H$ is rejected whenever sampled and is therefore covered by the Bernoulli full-step detection guarantee: if at least $M_{\mathrm{full}}$ predictable hard opportunities occur, the first $M_{\mathrm{full}}$ of them are all missed with probability $(1-p)^{M_{\mathrm{full}}}$, again for any history-adaptive strategy, and violations of the always-on structural invariants are rejected deterministically. This classification does not depend on how the GPU constructs its submitted gradient; a more elaborate forging strategy may change which class a step falls into, but not the classification itself. Throughout, ``accepted'' means that the run reaches certificate issuance after every deferred verification job has completed.

\paragraph{Normal FP32 path.}
Let $a_t$ denote the adversarial normalized deviation from the TEE FP32 reference on a body-path step, with
\begin{align*}
0\le a_t\le\tau_{\mathrm{abs}}.
\end{align*}
The adversary may choose $a_t$ adaptively from the previous protocol history, but must commit before the hidden current-step verification coin $V_t$ is revealed. Define
\begin{align*}
G_T=\sum_t a_t,
\qquad
Z_T=\sum_tV_ta_t.
\end{align*}
The cumulative check accepts only while $Z_T\le K_{\mathrm{sub}}$.

Using the martingale difference
\begin{align*}
X_t=(p-V_t)a_t
\end{align*}
and stopping when the cumulative deviation first reaches $G$, a Freedman--Bernstein bound gives
\begin{equation}
\begin{aligned}
& \Pr[
G_T\ge G,\,
Z_T\le K_{\mathrm{sub}}
]
\le \\
& \exp\!\left(
-
\frac{
(pG-K_{\mathrm{sub}})^2
}{
2p(1-p)\tau_{\mathrm{abs}}(G+\tau_{\mathrm{abs}})
+
\frac{2}{3}\tau_{\mathrm{abs}}(pG-K_{\mathrm{sub}})
}
\right).
\end{aligned}
\label{eq:g-sub-martingale}
\end{equation}
We define $G_{\mathrm{sub}}$ by numerically inverting Equation~\eqref{eq:g-sub-martingale} at failure probability $\beta_{\mathrm{sub}}$. Hence, cumulative body-path deviation exceeding $G_{\mathrm{sub}}$ evades this counter with probability at most $\beta_{\mathrm{sub}}$.

\paragraph{FP64 ambiguity path.}
Every accepted fallback use satisfies
\begin{align*}
a_t^{64}
=
\frac{
\Vert\bar g_t^{\mathrm{GPU}}-\bar g_t^{64}\Vert_2
}{C}
\le
\rho_{\mathrm{amb}}.
\end{align*}
Consider the first $M$ ambiguity opportunities generated by any predictable adversarial strategy, where whether the current submission constitutes such an opportunity is determined before the current verification coin is drawn. Because the verification coins are independent across steps and hidden until commitment, the audit indicators on these first $M$ opportunities are i.i.d.\ $\operatorname{Bernoulli}(p)$, so the number of audited ones satisfies
\begin{align*}
X_M\sim\operatorname{Binomial}(M,p).
\end{align*}
Define
\begin{align*}
M_{\beta_{\mathrm{amb}}}
=
\max\left\{
M:
F_{\mathrm{Bin}}(K_{\mathrm{amb}};M,p)
>
\beta_{\mathrm{amb}}
\right\}.
\end{align*}
Surviving beyond $M_{\beta_{\mathrm{amb}}}$ opportunities requires the accepted-escalation counter to stay at or below $K_{\mathrm{amb}}$ on the first $M_{\beta_{\mathrm{amb}}}+1$ of them, hence
\begin{align*}
\Pr\big[&\text{at least }M_{\beta_{\mathrm{amb}}}+1\text{ ambiguity opportunities occur}\\
&\text{and the counter stays}\le K_{\mathrm{amb}}\big]
\le
\beta_{\mathrm{amb}},
\end{align*}
for every history-adaptive strategy. The corresponding cumulative allowance is
\begin{align*}
G_{\mathrm{amb}}
=
\rho_{\mathrm{amb}}M_{\beta_{\mathrm{amb}}}.
\end{align*}

\paragraph{CIFAR-10 specialization.}
For CIFAR-10, which does not use FP64 adjudication, the same accounting applies with the ambiguity region defined directly in the FP32 metric:
\begin{align*}
\mathcal B&=\{t:z_t^{32}\le\tau_{\mathrm{abs}}\},\\
\mathcal A&=\{t:\tau_{\mathrm{abs}}<z_t^{32}\le\rho_{\mathrm{ill}}\},\\
\mathcal H&=\{t:z_t^{32}>\rho_{\mathrm{ill}}\}.
\end{align*}
Each accepted ambiguity opportunity contributes at most $\rho_{\mathrm{ill}}$, so $G_{\mathrm{amb}}=\rho_{\mathrm{ill}}M_{\beta_{\mathrm{amb}}}$ with the same Bernoulli-counter argument.

Combining the two channels,
\begin{align}
G_{\mathrm{extra}}
=
G_{\mathrm{sub}}
+
G_{\mathrm{amb}}.
\label{eq:extra-security-slack}
\end{align}
If the cumulative normalized deviation $G_{\mathrm{tol}}$ routed through the body and ambiguity channels exceeds this total, then at least one of the two component allowances is exceeded. Hence,
\begin{align*}
&\Pr[
G_{\mathrm{tol}}>G_{\mathrm{extra}}\ \text{and the run is accepted}
]\\
&\quad\le
\beta_{\mathrm{sub}}+\beta_{\mathrm{amb}}
=:
\beta_{\mathrm{extra}}.
\end{align*}

A submission with $z_t^{64}>\rho_{\mathrm{amb}}$ does not belong to $G_{\mathrm{extra}}$ and is rejected whenever sampled; the choice of reference on escalated steps is discussed under Status of the Guarantee below. The parameter table is given below.

\begin{table}[t]
\centering
\footnotesize
\setlength{\tabcolsep}{2pt}
\caption{Frozen verifier parameters per model family at the primary security operating point $p=0.1$; the numerical-tolerance guarantee is the pair $(G_{\mathrm{extra}},\beta_{\mathrm{extra}})$ with $\beta_{\mathrm{extra}}=\beta_{\mathrm{sub}}+\beta_{\mathrm{amb}}=0.05$ in every family. $z$-thresholds are in
clipping-norm units ($C{=}1$); $\tau_{\mathrm{num}}$ is stated in the
accountant's amplified units ($\times B/2C$) and scales with the batch size
($0.1565$ at $B{=}64$, $0.0783$ at $B{=}32$). CIFAR-10 replaces the fp64
layer with the same-metric hard cap $\rho_{\mathrm{ill}} = 2\tau_{\mathrm{abs}}$.
For the two false-abort rows see the notes below the table.}
\label{table:verifier-params}
\begin{tabular}{@{}lccc@{}}
\toprule
 & RoBERTa & GPT-2 & CIFAR-10 \\
\midrule
audit rate $p$ & 0.1 & 0.1 & 0.1 \\
$\tau_{\mathrm{abs}}$ & $1.60\times10^{-3}$ & $1.74\times10^{-3}$ & $7.30\times10^{-4}$ \\
$\rho_{\mathrm{amb}}$ / $\rho_{\mathrm{ill}}$ & $1.185\times10^{-2}$ & $4.891\times10^{-3}$ & $1.459\times10^{-3}$ \\
$\tau_{\mathrm{num}}$ & $1.5169$ & \shortstack{$0.1565$ /\\ $0.0783$} & --- \\
$K_{\mathrm{sub}}$ & 0.088 & 0.72 & 0.20 \\
$K_{\mathrm{amb}}$ & 25 & 17 & 5 \\
$(\beta_{\mathrm{sub}}, \beta_{\mathrm{amb}})$ & $(.005, .045)$ & $(.025, .025)$ & $(.025, .025)$ \\
aggregate-norm slack & $10^{-4}$ & $10^{-4}$ & $10^{-4}$ \\
worker pool $(K, S)$ & \shortstack{$(1,64)$ /\\ $(2,32)$} & $(2,32)$ & $(62{\times}1)$ \\
\midrule
$G_{\mathrm{extra}}$ & 5.478 & 9.496 & 2.511 \\
in-sample check (target $10^{-3}$) & \shortstack{$\le 2.8\times10^{-9}$\\ (3 traj.)} & $\le 8.7\times10^{-6}$ & $\le 1.1\times10^{-4}$ \\
held-out $q_{\mathrm{FA}}$ estimate & \shortstack{$\le 4\times10^{-100}$\\ (qqp-large replay)} & --- & --- \\
\bottomrule
\end{tabular}
\end{table}

\paragraph{Notes on Table~\ref{table:verifier-params}.} The two false-abort rows are evaluated under the Bernoulli protocol from the statistics observed as described in the sampling-laws paragraph. The held-out row is a trajectory-conditional upper bound from the exact statistics of the full-census replay and is the only false-abort estimate in this paper. The in-sample row is evaluated on the deployment-pair calibration trajectories: the RoBERTa entry covers three of the four RoBERTa trajectories (the qqp-large calibration trace being excluded as explained in the evaluation paragraph), the RoBERTa and GPT-2 entries are model-based upper values under the representativeness assumption for the systematic every-tenth-step sample, and the CIFAR-10 entry uses the rate model over the ${\approx}500$ checked steps of the $5{,}000$-step run; they certify only that the frozen parameters meet the target on the calibration data. Weight-decay convention for the LLM families (CIFAR-10 uses no weight decay): the released verifier and GPU trainer apply weight decay uniformly to every trainable parameter on both sides, matching the broad calibration campaign and the deployment-pair calibration trajectories (single-group AdamW). The overhead runs reported here were executed with the standard HF optimizer grouping on both sides, which exempts biases and LayerNorm weights from decay; the grouping changes the training trajectory but not the verifier implementation or worker configuration. The steering-attack experiments of Section~\ref{section:analysis_on_malicious_behavior} and Appendix~\ref{appendix:attack-adamw} also use the standard HF grouping.

\subsection{Empirical Steering-Scale Calibration}
\label{appendix:steering-equivalence}

This subsection is empirical. The formal guarantee is stated in the deviation
metric---$G_{\mathrm{extra}}$ bounds the cumulative normalized deviation accepted
through the tolerance channels (Equation~\eqref{eq:extra-security-slack})---and
does not depend on anything below. What follows supplies the unit conversion
behind the interpreted detection rate of Equation~\eqref{eq:effective-detection}:
we measure, on sampled steps, how much steering progress one normalized unit of
tolerated deviation buys relative to one full-power steering update. The
comparison is a \emph{same-state} one: both replacements are evaluated from
the same pre-step optimizer state, the tolerated set being a ball centered at
the honest aggregate and the full-power set the clipping ball. It is not a
post-center construction in which a full-power update is applied after the
trusted step.

For a tested step, let $\boldsymbol{g}^h$ denote the honest clipped aggregate the verifier
recomputes for that step. We compare two \emph{replacements} of $\boldsymbol{g}^h$ inside
the same training step, differing only in the set the forged gradient is drawn
from:
\begin{align*}
\text{tolerated:}\quad \lVert\boldsymbol{g}-\boldsymbol{g}^h\rVert_2\le rC,
\qquad
\text{full-power:}\quad \lVert\boldsymbol{g}\rVert_2\le C .
\end{align*}
The tolerated feasible set used here omits the always-on aggregate-norm
constraint $\lVert\boldsymbol{g}\rVert_2\le C(1+10^{-4})$ of the deployed
protocol and is therefore a relaxation of the deployed attacker's feasible
set; the comparison is consequently conservative in favour of the attacker.
At the tested states, even after adding the largest calibrated tolerance radius, the resulting norm remains far below the deployed cap, so the omitted constraint is inactive on these points; no general claim beyond the tested states is made.
Both are solved by the same steering optimizer from the same parameters and
optimizer state, and both are scored against the same realized DP noise, so the
comparison is a matched pair rather than two separate executions; neither is an
additional deployed step. Writing $P_t^{\mathrm{tol}}(r)$ and $P_t^{\mathrm{full}}$
for the progress each makes on the steering distance relative to the honest step,
define
\begin{align*}
R_t(r)
=
\frac{
P_t^{\mathrm{tol}}(r)
}{
rP_t^{\mathrm{full}}
},
\qquad
\widehat\kappa_{\mathrm{eq}}
=
\max_{(t,r)\in\mathcal T_{\mathrm{test}}}
R_t(r).
\end{align*}
Over nine RoBERTa checkpoints spanning $t/T=0.1$ to $0.9$, three GPT-2 states on E2E at $t/T=0.20$, $0.50$ and $0.90$, and
nine radii from $10^{-5}C$ to $C$, every measured $R_t(r)$ exceeds one by less than
$10^{-3}$, which is why the maximum rather than the mean is the quantity reported:
it is the direction that favours the adversary. We obtain
$\widehat\kappa_{\mathrm{eq}}^{\mathrm{RoBERTa}}=1.00064$ and
$\widehat\kappa_{\mathrm{eq}}^{\mathrm{LM}}=1.00038$. Thus, over the tested states and radii, a tolerated normalized radius $r$
has approximately $r$ times the steering effect of the corresponding
same-state full-power comparator; this same-state ratio is the conversion
used in Sections~\ref{sec:numerical-tolerance-security}
and~\ref{section:security_analysis}. The measurement covers the sampled steps,
radii, and model families of $\mathcal T_{\mathrm{test}}$, and the conversion in
Equation~\eqref{eq:effective-detection} further treats tolerated and full-power
deviations as additive and reads the same-state ratio as a step count; we
therefore report the interpreted full-step detection term as approximate, while the deviation bound of Equation~\eqref{eq:extra-security-slack}
holds irrespective of this calibration.

Two limits bound what this measures. The sub-threshold and ambiguity branches are
not independent evidence: they share one linearisation and their $R_t$ agree to
five decimals, so their agreement is arithmetic rather than corroboration. And the
matched noise draw is what makes the ratio tight --- it is the correct comparison
for a single step, since the noise is common to both branches and cancels along
the first-moment path, but under independent draws the same quantity varies by a
factor of two to three across seeds.

\paragraph{Transfer to the deployment trajectory family.} The conversion above is measured on the attack-evaluation trajectories, whose setup differs from the deployment trajectory family (Section~\ref{section:gpu-tee-discrepancy}). We use $\widehat\kappa_{\mathrm{eq}}$ only as an empirical scale when interpreting the deployed $G_{\mathrm{extra}}$ budget and do not claim it invariant across trajectory families; establishing that would require repeating this calibration on the deployment trajectories.

\subsection{Value-Aware Checking}
\label{appendix:value-aware}

The uniform protocol uses $p_t=p$ for every iteration. More generally, let
$p_t\in[0,1]$ be a hidden checking probability selected before training from
a public schedule satisfying the expected-work constraint
$\sum_{t=1}^{T}p_t=pT$. For a fixed set $\mathcal F$ of full malicious
deviations,
\begin{align*}
P_{\mathrm{det}}(\mathcal F)
=
1-\prod_{t\in\mathcal F}(1-p_t),
\end{align*}
so if certain training phases are known to provide larger attack value, the
verifier can assign larger $p_t$ to those phases while preserving the same
expected number of checks, and an adversary that concentrates its full
deviations on those phases faces a correspondingly higher detection
probability. This extension concerns the full-deviation detection channel
only. All numerical budgets $G_{\mathrm{sub}}$, $G_{\mathrm{amb}}$, and
$G_{\mathrm{extra}}$ reported in Table~\ref{table:verifier-params}, and all
experiments in this paper, use the uniform setting $p_t=p$; extending the
numerical-tolerance accounting to nonuniform $p_t$ is outside the present
evaluation.

\subsection{Status of the Guarantee}

$(G_{\mathrm{extra}},\beta_{\mathrm{extra}})$ is an analytic
high-probability security guarantee on the cumulative normalized deviation
routed through the numerical-tolerance channels, under the stated protocol
assumptions; $G_{\mathrm{extra}}$ alone is not a deterministic cap. The
guarantee is stated relative to the adaptive trusted reference selected by
the verifier on each step: it does not require the TEE FP32 and FP64
executions to define one canonical numerical trajectory, the difference
between these two trusted references is not itself charged as adversarial
deviation, and the bound concerns the GPU's deviation from the reference the
protocol uses to adjudicate that submission. The chain is calibration data
$\rightarrow$ frozen verifier parameters $\rightarrow$
$(G_{\mathrm{extra}},\beta_{\mathrm{extra}})$: the first arrow is empirical
parameter selection, the second is analytic security accounting, so a poor
estimate of the honest discrepancy distribution may make the chosen
parameters unsuitable for honest availability but does not invalidate the
bound evaluated at the parameters actually frozen.
Table~\ref{table:guarantee-hierarchy} summarizes the status of each
quantity used in this appendix.

\begin{table}[t]
\centering
\footnotesize
\caption{Status of the numerical-tolerance quantities.}
\label{table:guarantee-hierarchy}
\begin{tabular}{@{}p{0.30\columnwidth}p{0.24\columnwidth}p{0.38\columnwidth}@{}}
\toprule
Quantity & Status & Main dependency \\
\midrule
$1-(1-p)^{M_{\mathrm{full}}}$ & analytic & $M_{\mathrm{full}}$ full deviations under independent hidden Bernoulli checking \\
$(G_{\mathrm{extra}},\beta_{\mathrm{extra}})$ & analytic high-probability bound & Bernoulli protocol, frozen parameters, hidden coins \\
$q_{\mathrm{FA}}$ on the held-out census trajectory & trajectory-conditional estimate & exact trajectory, Bernoulli law; the only false-abort estimate \\
in-sample check from systematic samples & model-based calibration check & representativeness assumption; same trajectories as the calibration \\
future-run $q_{\mathrm{FA}}$ & empirical generalization & one held-out trajectory (qqp-large replay); otherwise untested \\
$G_{\mathrm{extra}}\to$ full-power steps & empirical interpretation & same-state steering-scale calibration, additive step-equivalent reading; measured on attack-experiment trajectories, transfer to the deployed pair assumed \\
Equation~\eqref{eq:effective-detection} & empirical interpretation & interpreted full-step detection term; not a formal theorem \\
\bottomrule
\end{tabular}
\end{table}

\section{Efficiency Analysis}
\label{appendix:efficiency_analysis}

The running time of our protocol consists of four main components: (1) GPU-side training time, including forward and backward propagation and clipping\footnote{While gradient clipping is a key component of DP-SGD, it is also commonly used in non-DP training to mitigate exploding gradients and improve training stability.}; (2) communication time between the GPU and the TEE; (3) TEE-side DP operations, such as noise generation and model update; and (4) TEE-side verification work incurred by probabilistic checking. We use $T_{\mathrm{train}}$, $T_{\mathrm{comm}}$, $T_{\mathrm{DP}}$, and $T_{\mathrm{veri}}$ to denote the corresponding total costs.

\paragraph{Communication and the main training path.} Under our communication-efficient split execution, the GPU sends only the clipped-and-averaged gradient to the TEE, rather than all per-example gradients. Let $S_G$ denote the size of this transmitted gradient. After the GPU-submitted gradient is committed, the TEE sends only a short random seed back to the GPU, which allows the GPU to reconstruct the same DP noise and maintain a synchronized local model and optimizer state.

Since the seed size is negligible compared with $S_G$, the steady-state communication cost per iteration can be approximated as
\begin{align*}
T_{\mathrm{comm}}^{\mathrm{iter}}
\approx
\frac{S_G}{\mathrm{Band}}
+
T_{\mathrm{sync}},
\end{align*}
where $\mathrm{Band}$ is the effective bandwidth of the GPU--TEE communication path and $T_{\mathrm{sync}}$ captures fixed synchronization costs such as message notification, TEE-boundary crossing, and buffer management. Since the transmitted gradient is an aggregate over the trainable parameters, the communication cost scales with the number of trainable parameters rather than with the batch size or the number of per-example gradients.

The GPU-side training, GPU--TEE communication, and TEE-side DP operations form the latency-critical execution path. Let
\begin{align*}
T_{\mathrm{iter}}
=
T_{\mathrm{train}}^{\mathrm{iter}}
+
T_{\mathrm{comm}}^{\mathrm{iter}}
+
T_{\mathrm{DP}}^{\mathrm{iter}}
\end{align*}
denote the average wall-clock time of this path for one iteration. For a run of $T$ iterations, the main training pipeline therefore requires approximately $N T_{\mathrm{iter}}$.

\paragraph{Hungry updating with deferred verification.} Under hungry updating, verification is removed from the latency-critical training path. Suppose each iteration is independently selected for checking with probability $p$, and let $T_{\mathrm{iter\text{-}veri}}$ denote the average work required to complete the full verification procedure for one checked iteration.

Assume that the TEE runs $n$ verification workers in parallel, in addition to the thread serving the main training path. On average, one verification task is generated every $1/p$ training iterations. Hence, the verification workers receive work at an average rate corresponding to
\begin{align*}
pT_{\mathrm{iter\text{-}veri}}
\end{align*}
units of verification work per training iteration. With $n$ workers, the verification system can process one check every $T_{\mathrm{iter\text{-}veri}}/n$ units of wall-clock time.

Therefore, the verification pipeline can keep pace with training when
\begin{align}
\frac{pT_{\mathrm{iter\text{-}veri}}}{n}
\le
T_{\mathrm{iter}},
\label{eq:verification-throughput-condition}
\end{align}
or equivalently,
\begin{align*}
n
\ge
\frac{
pT_{\mathrm{iter\text{-}veri}}
}{
T_{\mathrm{iter}}
}.
\end{align*}

When Equation~\eqref{eq:verification-throughput-condition} holds, verification work can be largely hidden behind the main training pipeline. Training may temporarily run ahead of verification, but a run is considered complete only after all verification tasks generated during training have finished successfully. Hence,
\begin{align*}
T_{\mathrm{total}}
=
T\,T_{\mathrm{iter}}
+
T_{\mathrm{drain}},
\end{align*}
where $T_{\mathrm{drain}}$ denotes the time required to drain any remaining verification tasks after the final training iteration. In the stable regime, $T_{\mathrm{drain}}$ is small, and therefore
\begin{align*}
T_{\mathrm{total}}
\approx
T\,T_{\mathrm{iter}}.
\end{align*}

On the other hand, if
\begin{align*}
\frac{pT_{\mathrm{iter\text{-}veri}}}{n}
>
T_{\mathrm{iter}},
\end{align*}
verification tasks are generated faster than the workers can process them, and a backlog accumulates. Under a steady-state approximation, the end-to-end running time becomes
\begin{align*}
T_{\mathrm{total}}
\approx
\max
\left\{
T\,T_{\mathrm{iter}},
\frac{
pT\,T_{\mathrm{iter\text{-}veri}}
}{n}
\right\}.
\end{align*}
In the verification-bottleneck regime, this reduces to
\begin{align*}
T_{\mathrm{total}}
\approx
\frac{
pT\,T_{\mathrm{iter\text{-}veri}}
}{n}.
\end{align*}
Thus, hungry updating converts verification from a synchronous per-check latency into a background throughput requirement. When sufficient verification parallelism is available, most verification work overlaps with the main training pipeline, resulting in only a small end-to-end overhead.

\section{MIA parameters}
\label{appendix:mia_parameters}

\begin{table}[!b]
\centering
\caption{Hyperparameters of the membership-inference experiments.}
\label{tab:mia_hyperparameters}
\scriptsize
\setlength{\tabcolsep}{2.5pt}
\begin{tabular}{@{}lc@{\hspace{0.8em}}lc@{}}
\toprule
\multicolumn{2}{c}{\textbf{IMIA}} & \multicolumn{2}{c}{\textbf{SHAPOOL}} \\
\cmidrule(lr){1-2}\cmidrule(lr){3-4}
Hyperparameter & Setting & Hyperparameter & Setting \\
\midrule
Number of imitative models & 10 & Number of shadow models & 10 \\
Imitative-out total epochs & 100 & Shadow pre-training epochs & 100 \\
Warm-up epochs & 80 & Shadow pre-training batch size & 128 \\
Imitation epochs & 20 & Shadow pre-training learning rate & 0.1 \\
Pivot fine-tuning epochs & 20 & Fine-tuning epochs & 3 \\
Shadow / imitation batch size & 256 & Fine-tuning batch size & 64 \\
Shadow / imitation learning rate & 0.1 & Fine-tuning learning rate & 0.1 \\
Shadow / imitation dropout & 0 & Number of experts & 5 \\
Pivot samples per class & 100 & MoE ratio & 0.5 \\
 &  & Dropout & 0 \\
\bottomrule
\end{tabular}
\end{table}
The parameters of the MIA experiments are shown in Table~\ref{tab:mia_hyperparameters}.

\end{document}